%% file: main.tex
\documentclass[letterpaper,journal]{IEEEtran}

\usepackage{amsmath,amsfonts}
\usepackage{amssymb}
\usepackage{algorithm}
\usepackage{algpseudocode}         
\usepackage{array}
\usepackage[caption=false,font=footnotesize,labelfont=sf,textfont=sf]{subfig}
\usepackage{textcomp}
\usepackage{stfloats}
\usepackage{url}
\usepackage{graphicx}
\usepackage{cite}

\usepackage{mathptmx}

\usepackage{booktabs}
\usepackage{xspace}

\usepackage[svgnames]{xcolor}

\definecolor{hillActive}{RGB}{31,119,180}   
\definecolor{hillPassive}{RGB}{44,160,44}   
\definecolor{hillTendon}{HTML}{C55A11}      

\usepackage{tikz}
\usetikzlibrary{positioning,arrows.meta,calc,backgrounds}

\edef\origsfdefault{\sfdefault}
\usepackage{carlito}
\edef\carlitofamily{\sfdefault}
\let\sfdefault\origsfdefault
\newcommand{\figsf}{\fontfamily{\carlitofamily}\selectfont}

\graphicspath{{./}{imgs/}{../imgs/}}

\DeclareGraphicsExtensions{.pdf,.png,.jpg,.jpeg}

\PassOptionsToPackage{hyphens}{url}

\usepackage[hyperfootnotes=false,breaklinks=true,colorlinks=false,
            pdfborder={0 0 0}]{hyperref}
\usepackage{orcidlink}


\begin{document}

\bstctlcite{IEEEexample:BSTcontrol}

\title{RLMuscle: Muscle-Driven Character Animation\\ via Reinforcement Learning}

\author{%
  Chunlei~Li\,\orcidlink{0000-0002-9822-9465},
  Siyuan~Yu\,\orcidlink{0009-0003-1310-0094},
  Yang~Gao\,\orcidlink{0000-0002-9149-3554},
  Shuai~Li\,\orcidlink{0000-0003-4182-1588},
  Peng~Yu\,\orcidlink{0000-0002-8652-2744},
  Aimin~Hao\,\orcidlink{0000-0002-5774-6706},
  and~He~Wang\,\orcidlink{0000-0002-2281-5679}
\thanks{Chunlei~Li, Yang~Gao, Shuai~Li, Peng~Yu, and Aimin~Hao are with Beihang University, Beijing, China.}
\thanks{Chunlei~Li's primary affiliation is Beihang University, China. This work was carried out during his academic visit to University College London, UK.}
\thanks{Siyuan~Yu is with Westlake University, Hangzhou, China.}
\thanks{He~Wang is with University College London, London, UK. He is the corresponding author (e-mail: realcrane@gmail.com).}}

\markboth{Preprint}%
{Li \MakeLowercase{\textit{et al.}}: RLMuscle: Muscle-Driven Character Animation via Reinforcement Learning}

\maketitle

\begin{abstract}
Body deformation caused by muscle contraction plays a vital role in the visual realism of character animation, yet in most 3D animation pipelines artists have to manually specify muscle activations, which is time-consuming and gives no guarantee of physical realism. To address this, we propose a new method in which a reinforcement learning (RL) controller discovers per-muscle activations that track reference motions without manual tuning. Existing RL muscle controllers stop at 1D muscles; we go one step further, keeping the 1D model inside the RL control loop as a fast surrogate, and using a 3D volumetric simulation to produce the final deformation. Because both stages share the very same constitutive law, the muscle deformation is produced by the same activations that generated the motion. We further add a measured electromyography (EMG) prior that steers the redundant activation solution toward real human muscle recruitment. To achieve stable, efficient, and realistic simulation, we propose a novel geometric multigrid for Extended Position-Based Dynamics (XPBD) that is built on a coarse tetrahedral cage with barycentric-coordinate prolongation and closes a V-cycle on tetrahedral meshes. Within this GPU-parallelized solver, we inject the same Hill-type force--length law as an energy-based fiber constraint that turns per-fiber activation into realistic muscle contraction and bulging. We demonstrate volumetric muscle simulation on a 194-muscle human body with a diverse set of motions (walk, kick, swim, overhead-squat, etc.), simulating the full-body model at 18\,ms per frame on a single RTX~4090. Our work takes a step toward character animation that is not only visually convincing but also biomechanically grounded. The website with code and videos is at \url{https://chunleili.github.io/project-page-rlmuscle/}.
\end{abstract}

\begin{IEEEkeywords}
Muscle simulation, XPBD, reinforcement learning, physics-based animation, character animation.
\end{IEEEkeywords}



\input{main-body}


\bibliographystyle{IEEEtran}
\bibliography{refs}

\end{document}

%% file: main-body.tex
\section{Introduction}
\label{sec:intro}

\IEEEPARstart{I}{n} animation, film, games, and virtual reality (VR), realistic muscle behavior is essential for believable character animation. Yet production pipelines still rely on \emph{hand-authored} muscle activations, produced by one of two major authoring methods. The first is a \emph{tension line}: the artist binds a line onto the skeleton, and the muscle associated with that line is activated once the line contracts below a set length threshold, so activation is triggered geometrically by bone motion. The second is \emph{keyframing}: the artist directly animates each muscle's activation value over time as a keyframed curve. Both encode activations chosen for visual plausibility, not physical plausibility. Consequently, there is no guarantee that hand-crafted tension can produce physically valid motions, because artists effectively decouple muscle deformation from motion dynamics. This expertise-driven workflow is not only labor-intensive; it also cannot answer the question at the heart of physics-based animation: \emph{what muscle activations actually produce a given motion?} This is the \emph{muscle inverse problem}: recovering the hidden control signals from observed motion. A principled solution would replace hand-tuning with physically consistent activations derived from the target motion, so that the resulting muscle deformation reflects the underlying dynamics rather than a visually plausible guess.

On the other hand, recent works on muscle simulation driven by reinforcement learning (RL)~\cite{lee_scalable_2019,myosuite,feng_musclevae_2023,park_magnet_2025,li_towards_2026} are mainly limited to a simplified 1D Hill-type muscle model, which falls short of the 3D volumetric muscle deformation needed for visual realism. The muscle inverse problem is challenging because the map from activation to joint motion is highly nonlinear, and the control space is high-dimensional (hundreds of muscles), which leads to intractability of analytical inversion, and extremely low sample efficiency in RL training.

We solve the muscle inverse problem by \emph{forward dynamics}, i.e.\ searching for the activations that, once simulated, reproduce a target motion. This poses two challenges:
\begin{enumerate}
    \item \textbf{Control.} The map from muscle activation to joint motion is highly underdetermined and nonlinear, so analytical inversion is intractable, and naive RL exploration in this space is sample-inefficient.
    \item \textbf{Physics.} The numerical solver must handle volumetric muscle that is nonlinear, anisotropic, and nearly incompressible while remaining both stable and efficient.
\end{enumerate}

We present the first system that addresses both challenges. Given a reference motion (motion capture or keyframed animation) and rest-pose 3D muscle/skeleton assets as input, it outputs a skeletal--muscular animation driven by physically consistent activations. Our work makes three technical contributions.

\textbf{Technical contribution~1: a muscle-control pipeline that decouples 1D control from 3D simulation under a unified constitutive law for fibers,} solving the muscle inverse problem and freeing artists from hand-tuning (addressing Control). Existing RL muscle controllers~\cite{lee_scalable_2019, feng_musclevae_2023, park_magnet_2025, li_towards_2026, myosuite} actuate 1D Hill-type muscles only, so they recover joint motion but not the 3D volumetric deformation that visual realism demands; conversely, optimization-based methods~\cite{lee_dexterous_2018} that place the volumetric solver \emph{inside} the control loop are unstable and did not converge in our tests. We propose to use the 1D model \textbf{as a surrogate} in RL and to produce the deformation from the 3D model. To bridge the gap between the two, we introduce a \textbf{new fiber constraint based on the Hill--Millard constitutive curve}, which is used in both the 1D and the 3D model. Following MASS~\cite{lee_scalable_2019}, a \emph{policy network} and a \emph{muscle network} are \emph{co-trained} in a single PPO~\cite{schulman_proximal_2017} loop against a motion-mimicking reward~\cite{peng_deepmimic_2018} on the 1D surrogate. The recovered activations then drive the 3D volumetric muscles through the \emph{same} Hill-type constitutive law~\cite{millard_flexing_2013} used by the surrogate, in a single forward pass with the 3D solver kept \emph{out} of the control loop, so the volumetric deformation is consistent with the dynamics that produced the motion, reproducing realistic muscle bulging, stiffening, and contraction rather than a visually plausible guess.

\textbf{Technical contribution~2: a physiological electromyography (EMG) prior} that steers the redundant activation solution toward measured human muscle recruitment (addressing Control). The activation-to-torque map is redundant, and existing muscle networks~\cite{lee_scalable_2019, feng_musclevae_2023, park_magnet_2025, li_towards_2026, myosuite} reduce that redundancy from torque and effort alone, with nothing tying the result to a real recruitment pattern, so their activations drift away from measured human EMG. We propose to inject a measured surface EMG envelope~\cite{moreira2021lower} into the muscle-network loss, matched per synergist group and folded onto a common gait cycle, which pulls the recovered activations back toward the measured envelope (\S\ref{sec:res-emg}).

\textbf{Technical contribution~3: a Petrov--Galerkin geometric multigrid for volumetric XPBD,} integrated into a GPU-parallel framework that runs the full-body volumetric muscle model at 18\,ms per frame on a single RTX~4090 (addressing Physics). Previous position-based hierarchies, such as HPBD~\cite{Muller2008-HPBD}, are formulated for cloth rather than tetrahedral meshes and never close a complete V-cycle, so they act only as a low-resolution driver; algebraic constructions~\cite{Li2025-MGPBD} instead assemble a global matrix and thereby give up XPBD's local, per-constraint solve. We propose to coarsen a tetrahedral cage geometrically, prolongate by barycentric coordinates, and close a full V-cycle with a Petrov--Galerkin restriction, which keeps the high-stiffness solve stable and efficient. The framework further contributes muscle--bone collision handling based on a signed distance field (SDF).

\begin{figure}[!t]
  \centering
  \includegraphics[width=\linewidth]{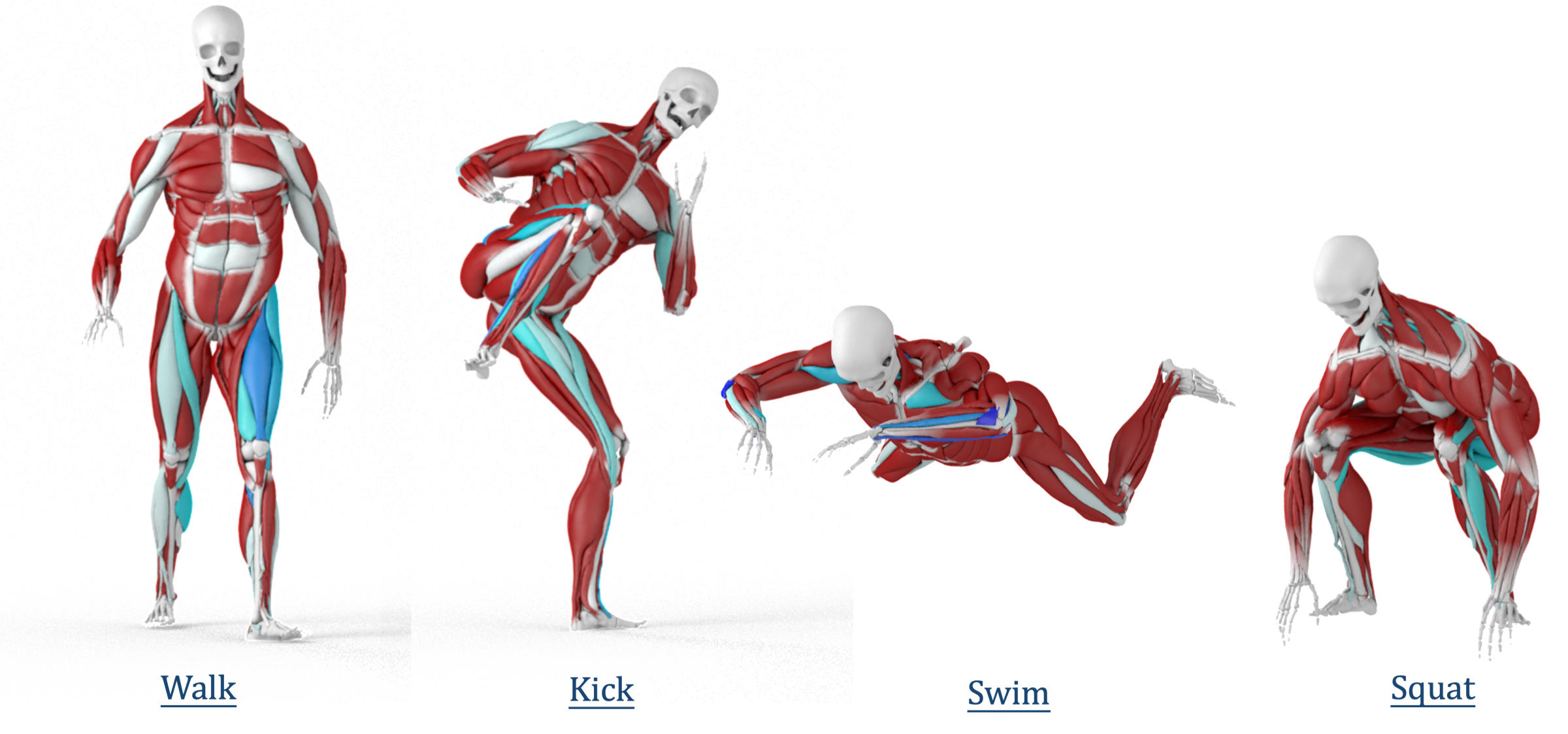}
  \caption{Automatically activated muscle-driven character animation for diverse motions: (a)~walking, (b)~kicking, (c)~swimming, and (d)~squatting. Blue intensity shows the activation level; different shades of red distinguish muscle tendons from muscle bellies.}
  \label{fig:teaser}
\end{figure}

\section{Related Work}
\label{sec:related}

\textbf{Position-Based Dynamics.} PBD~\cite{muller_position_2007} is a popular framework for real-time deformable object simulation. It is widely used in applications such as PhysX~\cite{physx} and Houdini Vellum~\cite{Houdini}. Extended PBD (XPBD)~\cite{macklin_xpbd_2016} reformulates PBD as a dual-space energy minimization framework and introduces a compliance term to transform the infinite-stiffness to a finite-stiffness problem, which improves the convergence. Bender et al.~\cite{bender_position-based_2014} cast Finite Element Method (FEM) strain energy as position-based constraints, which makes it possible to apply various FEM-based constitutive laws in PBD. We refer readers to the survey~\cite{Bender2017-Survey-PBD} for more details. We use XPBD as our simulation engine and propose a novel muscle fiber constraint tied to the physiological properties of muscle.

\textbf{Multigrid.} Multigrid~\cite{Briggs2000-MgTut} is a classical numerical method for accelerating and stabilizing simulation, running an iterative solver on grids of different resolutions to eliminate low-frequency errors. Unlike grid-based fluid solvers, which have a natural hierarchical structure, soft bodies provide no such structure, so building a hierarchy on them is challenging. Hierarchical PBD (HPBD)~\cite{Muller2008-HPBD} introduces a hierarchical solver for PBD simulated cloth, whose coarsening is based on edge collapsing. For volumetric solids, by contrast, the hierarchy is built by coarsening the simulation mesh itself, and how easily this can be done depends on the element type: coarsening is straightforward and efficient on hexahedral meshes~\cite{zhu2010-Multigrid-Solid}, but the tetrahedral meshes prevalent in computer graphics are harder to coarsen. Several strategies address this: some build a multilevel XPBD through gradual rigidifying~\cite{Mercier2024-Rigidifying}, while barycentric-coordinate prolongation~\cite{Georgii-Westermann2006-multigrid} constructs a hierarchy directly on tetrahedral meshes and has proved successful in FEM. Our previous work, Multi-Grid Position-Based Dynamics (MGPBD)~\cite{Li2025-MGPBD}, utilized algebraic multigrid to build the hierarchy. However, algebraic multigrid requires assembling a global matrix, which sacrifices the flexibility and lightweight nature of XPBD's local, per-constraint solve; moreover, building the hierarchy purely from the numerical matrix lacks geometric and physical intuition. In this work we propose a novel barycentric-coordinate-based geometric multigrid for XPBD, which accelerates the convergence of the stiff volume and incompressibility constraints and yields more stable and efficient volumetric simulation than single-level XPBD.

\textbf{Volumetric muscle simulation.}
Muscle is a special type of soft tissue, characterized by anisotropic fiber forces and strong incompressibility. Kim et al.~\cite{kim_dynamic_2020, kim_anisotropic_2019, smith_stable_2018} proposed energy potentials for FEM that yield an inversion-free muscle material matching the constitutive law of real muscle. Angles et al.~\cite{angles_viper_2019} used position-based elastic rods with volumetric scale to model muscle bundles.

\textbf{Muscle-actuated Hill muscle control by RL.}
In recent years, many researchers have used 1D Hill-type muscles to physically drive the dynamics of human motion (activation $\rightarrow$ muscle force $\rightarrow$ bone motion, i.e.\ forward dynamics). The Hill-type model approximates each muscle as a polyline anchored to local positions on the bones, and reinforcement learning is commonly employed to train a neural network that finds the activations producing the desired motion. OpenSim~\cite{dembia_opensim_2020} provides 1D musculotendon simulation, and Myosuite~\cite{myosuite} wraps it in an RL framework for muscle-actuated control. The most prominent of these is MASS (Muscle-Actuated Skeletal System) by Lee et al.~\cite{lee_scalable_2019}, a two-level imitation-learning architecture that actuates 346 muscles; we adopt its dual-network design (\S\ref{sec:rl}). A line of follow-ups inherits the same control formulation: MuscleVAE~\cite{feng_musclevae_2023}, MAGNET~\cite{park_magnet_2025}, and MuscleMimic~\cite{li_towards_2026}, alongside torque-actuated trackers such as DeepMimic~\cite{peng_mimicKit_2025,peng_deepmimic_2018} with Proximal Policy Optimization (PPO)~\cite{schulman_proximal_2017}. These methods all drive only \emph{1D line-segment} Hill muscles; building on this 1D Hill-type RL control, we add a 3D volumetric muscle simulation driven by the \emph{same} Hill law, turning the biomechanics into visually compelling muscle deformation.

\textbf{Inverse muscle activation for volumetric muscle.}
Some works also address the inverse muscle activation problem with 3D volumetric deformation, similar to ours, but they recover the activations by static optimization rather than inside a forward-dynamic control loop. Han et al.~\cite{han_neural_2024} solve a per-pose torque balance as a quadratic program (QP) and train an ML-deformer~\cite{Bailey:2018:FDD} to reproduce muscle deformation from joint state and activation, but such a static solve cannot guarantee that the recovered activations actually drive the target motion (activation $\rightarrow$ muscle force $\rightarrow$ bone motion). Lee et al.~\cite{lee_dexterous_2018} embed Hill-type muscles in Projective Dynamics and obtain activations from a per-frame QP, which runs the volumetric solver \emph{inside} the trajectory-optimization control loop, but in our tests it was very unstable and never converged. We instead learn the activations \emph{inside} the RL control loop on a 1D Hill-type surrogate, so they are exactly the forces that drive the motion (\S\ref{sec:rl}), making the deformation genuinely physically actuated rather than fitted or pre-scripted.

\section{Method}
\label{sec:method}

Our system couples a 1D Hill-type muscle model ($N_m=284$) with an RL controller, then uses a volumetric muscle model ($194$ muscles) with the same Hill-type constitutive law to simulate the deformation. We first outline the pipeline (\S\ref{sec:pipeline}) and the RL co-training that solves the inverse problem (\S\ref{sec:rl}), then detail the 1D Hill dynamics (\S\ref{sec:physics}), the 3D volumetric simulation (\S\ref{sec:xpbd}), and the motion retargeting and muscle mapping that bridge the 1D and 3D models (\S\ref{sec:retarget}).

\subsection{Overview}
\label{sec:pipeline}

\begin{figure}[t]
  \centering
  \includegraphics*[width=\linewidth]{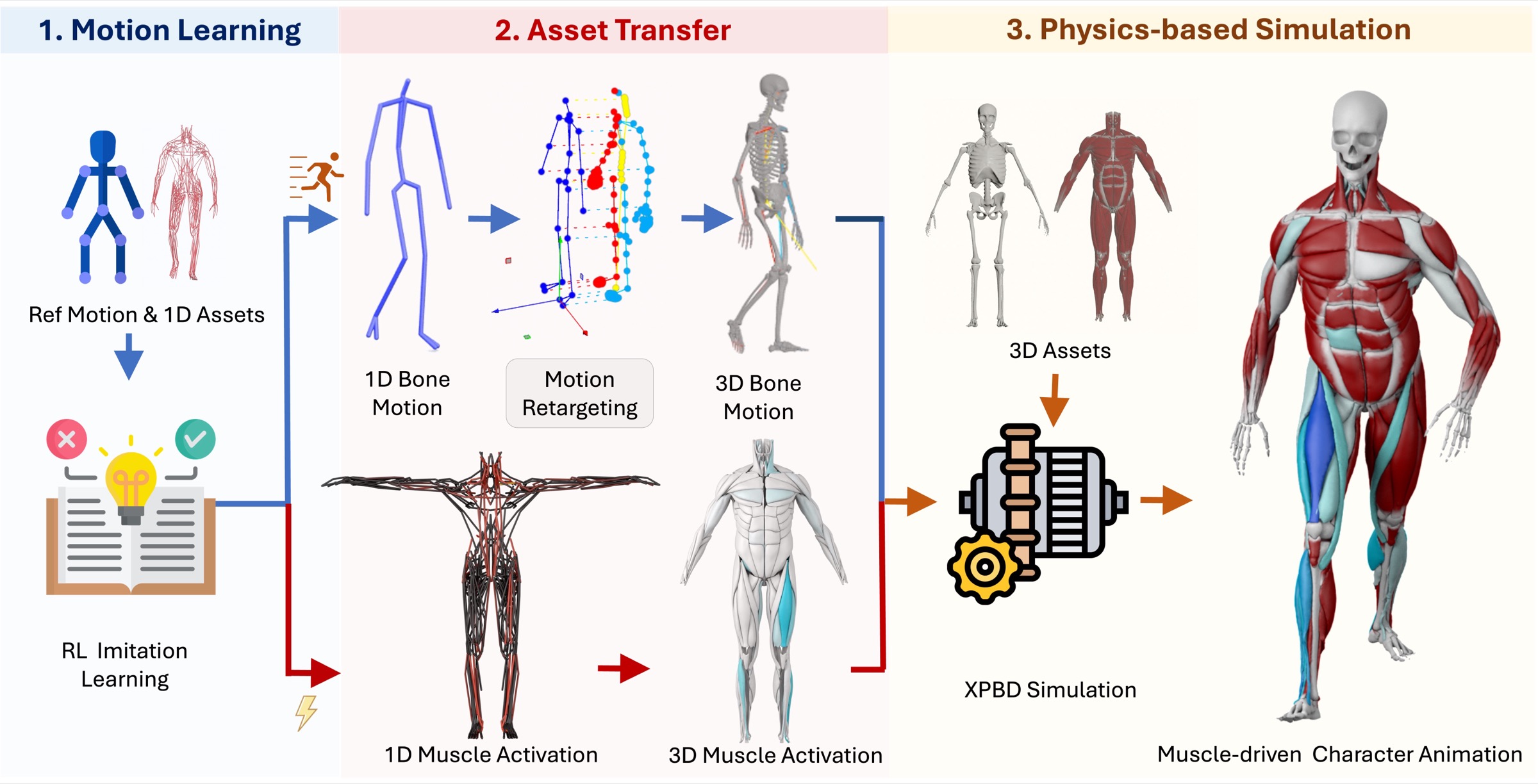}
  \caption{Pipeline overview. RL imitation learning runs entirely on the 1D asset; retargeting and muscle mapping then lift its output to a 3D skeleton and per-fiber activations, which the XPBD engine simulates in a single forward pass.}
  \label{fig:pipeline}
\end{figure}

\begin{figure*}[htbp]
    \centering
    \resizebox{\textwidth}{!}{%
    \begin{tikzpicture}[
        font=\figsf\small,
        >={Stealth[length=3.2mm, width=2.6mm]},
        ic/.style={inner sep=0pt},
        capt/.style={font=\figsf\footnotesize, align=center, inner sep=1.5pt, text=black!80},
        acbox/.style={rounded corners=3pt, draw=black!45, line width=0.7pt,
            fill=black!2, inner xsep=4pt, inner ysep=4pt, align=left},
        flow/.style={->, line width=1.5pt, black!75},
        musf/.style={->, line width=1.5pt, red!72!black},
        bonef/.style={->, line width=1.5pt, blue!62!black},
        upd/.style={->, line width=1.2pt, red!70!black, dashed},
        fbk/.style={->, line width=1.2pt, black!60, dashed},
        outf/.style={->, line width=0.9pt},
        lbl/.style={font=\figsf\footnotesize, inner sep=1.5pt}
    ]
    \def\isz{10mm}
    \node[ic] (st)   at (-0.15,0)  {\includegraphics[height=\isz]{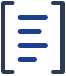}};
    \node[acbox] (ac) at (2.5,0) {%
        \raisebox{-2.6mm}{\includegraphics[width=9mm]{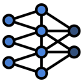}}\,policy NN $\pi_\phi$\\[1.6mm]
        \raisebox{-2.6mm}{\includegraphics[width=9mm]{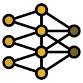}}\,value NN $V_\psi$};
    \node[ic] (pd)   at (5.1,0)  {\includegraphics[height=\isz]{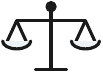}};
    \node[ic] (mus)  at (7.2,0)  {\includegraphics[width=11mm]{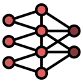}};
    \node[ic] (hill) at (9.9,0)  {\includegraphics[height=\isz]{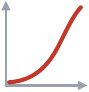}};
    \node[ic] (mjc)  at (12.0,0) {\includegraphics[height=\isz]{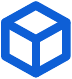}};
    \node[ic] (mot)  at (13.6,0) {\includegraphics[height=\isz]{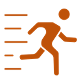}};
    \node[capt, below=0.3mm of st]   {state $\mathbf{s}_t$};
    \node[capt, below=0.3mm of pd]   {Stable PD};
    \node[capt, above=0.3mm of mus, xshift=-2mm]  {muscle NN $\mu_\theta$};
    \node[capt, below=0.3mm of hill] {Hill};
    \node[capt, below=0.3mm of mjc]  {MuJoCo};
    \node[capt, above=0.3mm of mot, text=blue!62!black]  {motion};
    \node[ic] (ppo) at (2.5,-2.05) {\includegraphics[height=\isz]{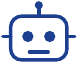}};
    \node[draw=red!75!black, line width=0.8pt, rounded corners=3pt, fill=white,
          inner sep=3pt, font=\figsf\footnotesize] (mse) at (7.2,-1.45)
        {\raisebox{-1.3mm}{\includegraphics[height=4.5mm]{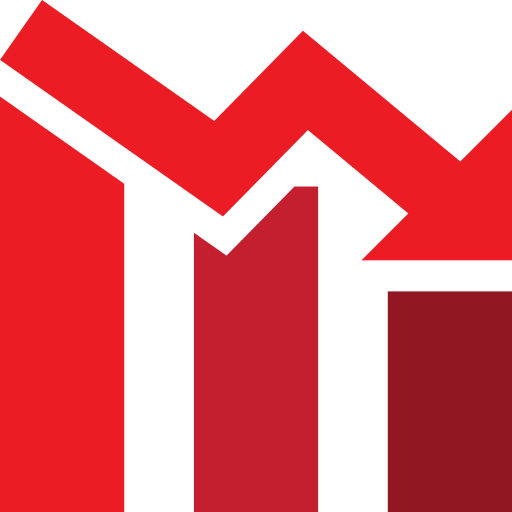}}\,%
         $L_{\mu}$};
    \node[ic] (tgt) at (13.6,-2.05) {\includegraphics[height=\isz]{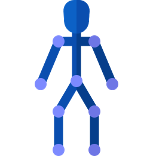}};
    \node[ic] (rew) at (11.6,-2.05) {\includegraphics[height=\isz]{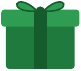}};
    \node[capt, below=0.3mm of ppo] {PPO};
    \node[capt, below=0.3mm of tgt] {target (ref.\ motion)};
    \node[capt, below=0.3mm of rew] {reward};

    \draw[flow] (st) -- (ac);
    \draw[flow] (ac)   -- (pd)   node[lbl, midway, above] {$\Delta\boldsymbol{\theta}_t$};
    \draw[flow] (pd)   -- (mus)  node[lbl, midway, above] {$\boldsymbol{\tau}^{\text{des}}$};
    \draw[musf] (mus)  -- (hill);
    \node[ic] (bolt) at (8.25,0.32)
        {\includegraphics[height=5mm]{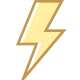}};
    \node[lbl, text=red!72!black, right=0.4mm of bolt] {$\mathbf{a}^{\mathrm{M}}$};
    \draw[outf, red!72!black] (8.25,0.62) -- (8.25,1.08);
    \node[lbl, text=red!72!black] at (8.25,1.25) {output};
    \draw[flow] (hill) -- (mjc)  node[lbl, midway, above] {$\boldsymbol{\tau}^{\text{musc}}$};
    \draw[bonef] (mjc)  -- (mot);
    \draw[outf, blue!62!black] (12.75,0.15) -- (12.75,1.08);
    \node[lbl, text=blue!62!black] at (12.75,1.25) {output};

    \coordinate (tmout) at ($(hill.east)!0.5!(mjc.west)$);
    \draw[flow] ([yshift=-5mm]pd.south) |- (mse.west);
    \draw[flow] (tmout)     |- (mse.east);
    \draw[upd]  (mus.south |- mse.north) -- (mus.south) node[lbl, pos=0.35, right] {update $\mu_\theta$};

    \draw[flow] (mot.south) -- ++(0,-0.65) -| (rew.north);
    \draw[flow] (tgt) -- (rew);
    \draw[flow] (rew.west) -- (ppo.east)
        node[lbl, midway, below] {$r_t$ (tracking reward)};
    \draw[upd]  (ppo) -- (ac) node[lbl, pos=0.32, right] {update $\phi,\psi$};
    \draw[fbk] (mjc.north) -- ++(0,1.0) -| (st.north)
        node[lbl, pos=0.15, above] {$\mathbf{s}_{t+1}$};
    \end{tikzpicture}%
    }%
    \caption{RL pipeline, i.e.\ the \emph{RL imitation learning} stage of Fig.~\ref{fig:pipeline} opened up. A per-joint PD-target offset $\Delta\boldsymbol{\theta}_t$ is mapped through Stable~PD and the muscle network to $N_m{=}284$ activations $\mathbf{a}^{\mathrm{M}}\!\in\![0,1]^{N_m}$. Both losses (dashed red) are evaluated each iteration, so $\pi_\phi$ and $\mu_\theta$ are co-trained in a single PPO loop rather than learned sequentially. There are two outputs, corresponding to Fig.~\ref{fig:pipeline}. The first is the \emph{1D Bone Motion} (blue), the second is the \emph{1D Muscle Activation} $\mathbf{a}^{\mathrm{M}}$ (red, marked by the bolt).}
    \label{fig:rl-pipeline}
\end{figure*}
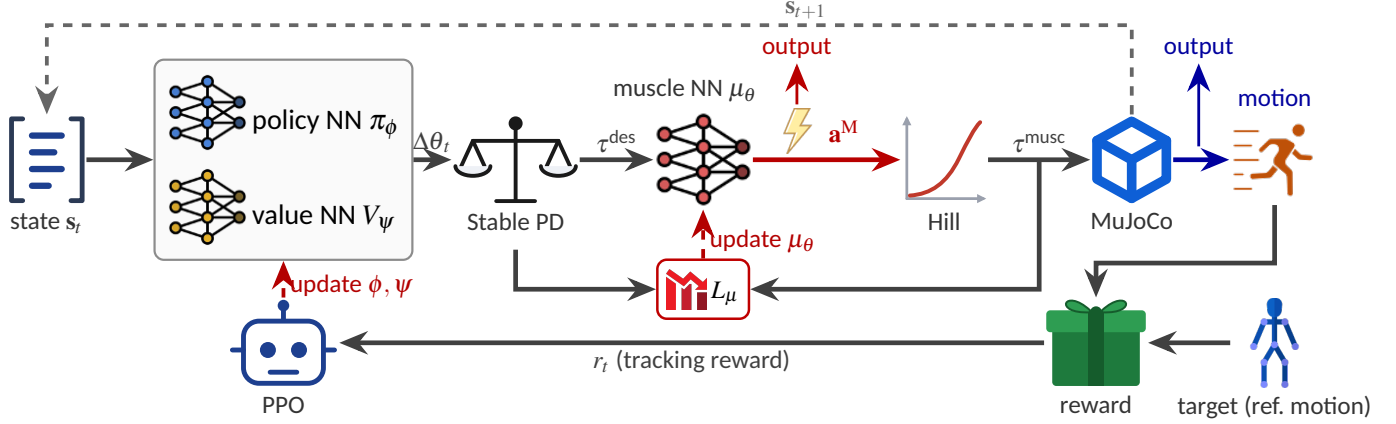

Figure~\ref{fig:pipeline} summarizes the system. The input is a 1D musculoskeletal asset together with a reference motion clip in Biovision Hierarchy (BVH) format. A reinforcement-learning \emph{imitation-learning} stage solves the muscle inverse problem on this 1D asset (284 muscles), jointly producing the muscle-driven bone motion and the corresponding 1D \emph{muscle activation trajectory}. A \emph{motion-retargeting} step transfers the bone motion onto a 3D skeleton, while a \emph{muscle-mapping} step lifts the per-musculotendon 1D activations to 3D activations. Both streams and 3D assets (bone and muscle geometries) are then fed to the XPBD-based 3D \emph{simulation engine} (\S\ref{sec:xpbd}), which produces the volumetric muscle deformation used for the final 3D rendering. The 1D track therefore acts as a fast surrogate that solves the hard inverse problem once; the 3D track is a single forward simulation at inference. The RL co-training that solves the inverse problem (Fig.~\ref{fig:rl-pipeline}, Alg.~\ref{alg:train}) is detailed in \S\ref{sec:rl}.

\subsection{Reinforcement Learning Control}
\label{sec:rl}

The RL stage solves \emph{control} challenges: it discovers the muscle activations ($N_m=284$) that make the simulated character track a reference motion. We split control across a \emph{policy network} $\pi_\phi$ and a \emph{muscle network} $\mu_\theta$~\cite{lee_scalable_2019} and \emph{co-train} them inside a single PPO loop against the motion-tracking reward (Fig.~\ref{fig:rl-pipeline}, Alg.~\ref{alg:train}) rather than learning them sequentially, so each rollout supplies a gradient for both and they adapt to each other. A third, training-only \emph{value network} $V_\psi$ serves as PPO's critic but never participates in control.

\subsubsection{Two-Network Co-Training}

We train the controller for muscle-driven motion tracking with PPO, summarized in Alg.~\ref{alg:train}. Letting the policy explore the $N_m{=}284$ muscle activations directly is sample-inefficient, so the policy instead acts in the compact space of per-joint PD-target offsets: Stable proportional--derivative (PD) control~\cite{tan_stable_2011} turns its action into a desired joint torque $\boldsymbol{\tau}^{\text{des}}\!\in\!\mathbb{R}^{N_\theta}$ ($N_\theta{=}50$ degrees of freedom, DOF), the muscle network maps that torque to the activations $\mathbf{a}^{\mathrm{M}}_t\!\in\![0,1]^{N_m}$ (M for muscles), and the Hill model realizes the muscle torque that drives MuJoCo~\cite{mujoco}.  Both networks learn from the same rollout, PPO updating $\pi_\phi, V_\psi$ and a mean squared error (MSE) loss updating $\mu_\theta$. The observation is $\mathbf{s}_t = [\, \mathbf{b}_t,\; \mathbf{v}_t,\; \phi_t \,]$, 
where $\mathbf{b}_t\!\in\!\mathbb{R}^{66}$ stacks the center-of-mass position of each of the $22$ non-root bodies in the root frame, $\mathbf{v}_t\!\in\!\mathbb{R}^{69}$ the world-frame center-of-mass velocities of all $23$ bodies, and $\phi_t\!\in\![0,1)$ is the phase of the reference clip ($\dim(\mathbf{s}_t){=}136$). 

\subsubsection{Policy Network}

\textbf{Motion-mimicking reward.} Following DeepMimic~\cite{peng_deepmimic_2018} we use an exponential-kernel reward composed of pose, velocity, and end-effector terms, augmented with whole-body center of mass (COM) and pelvis-orientation terms for balance-critical motions; the pose error is the $SO(3)$ geodesic between current and target joint frames:
\begin{align}
  \mathbf{p}_j      &= \log\!\big(R_{j}^{\top}\, R^{*}_{j}\big)^{\!\vee} \in \mathbb{R}^3, \label{eq:logmap}\\
  r_{\text{pose}}   &= \exp\!\Big({-}k_p \textstyle\sum_{j\in\mathcal{J}}\, \|\mathbf{p}_j\|^2\Big), \label{eq:reward_pose}
\end{align}
\vspace{-1.4em}
\begin{equation}
  r_t = r_{\text{ee}} \cdot \big(w_p\, r_{\text{pose}} + w_v\, r_{\text{vel}} + w_c\, r_{\text{com}} + w_u\, r_{\text{up}}\big),
  \label{eq:reward}
\end{equation}
where $r_t$ is the tracking reward; $R_j, R^*_j \in SO(3)$ are the current/target rotation matrices of joint $j$; $\log(\cdot)^{\!\vee}$ is the matrix log map returning the axis--angle vector; $\mathcal{J}$ is the set of tracked joints; and asterisks mark reference-motion targets. The remaining terms share the same kernel $\exp({-}k\,\lVert\cdot\rVert^2)$ applied to their own error: joint velocity for $r_{\text{vel}}$, the positions $\mathbf{x}_k$ of the five end-effector links for $r_{\text{ee}}$, the whole-body center of mass $\mathbf{c}$ for $r_{\text{com}}$, and the pelvis up axis $\mathbf{u}{=}R_{\text{pelvis}}\mathbf{e}_y$ for $r_{\text{up}}$, which gives a yaw-free torso-tilt error (full expressions in the supplementary). End-effector errors are taken in the COM-relative frame so root drift does not corrupt limb-placement credit, and $r_{\text{ee}}$ enters multiplicatively so gross limb misplacement suppresses local joint credit. The exponential kernel confines each term to $[0,1]$ and saturates smoothly as the tracking error grows. Balance-critical motions (squat, single-leg kick) shift the task weights $(w_p,w_v,w_c,w_u)$ toward the COM and pelvis-up terms and add the root pose to $\mathcal{J}$; without these terms the policy matches the pose but loses balance and falls. Kernel widths, per-task weights and the Mixamo~\cite{mixamo} clips are listed in the supplementary.



\begin{algorithm}[t]
\caption{RL Policy Training (Co-training PPO)}
\label{alg:train}
\begin{algorithmic}[1]
\Require Reference motion $\mathcal{D}$, muscle--skeleton model
\State Initialize policy $\pi_\phi$, value $V_\psi$, muscle-NN $\mu_\theta$, replay buffer $\mathbf{B}$, muscle torque buffer $\mathbf{B}^\text{musc}$
\For{iteration $=1,2,\dots$}
    \State \textbf{// Collect Trajectories}
    \For{$t=0,\dots,T-1$ with $N_{\text{env}}$ envs in parallel}
        \State $\mathbf{a}_t \sim \pi_\phi(\mathbf{s}_t)$ \Comment{action from policy-NN}
        \State $\boldsymbol{\tau}^{\text{des}}_t \gets \textsc{Stable PD}(\mathbf{s}_t,\mathbf{a}_t)$ \Comment{PD control}

        \State $\mathbf{a}^{\mathrm{M}}_t \gets \mu_\theta(\mathbf{s}_t,\boldsymbol{\tau}^{\text{des}}_t)$ \Comment{muscle-NN: $\boldsymbol{\tau}^{\text{des}}\!\to\!\mathbf{a}^{\mathrm{M}}$}
        \State $\boldsymbol{\tau}^{\text{musc}}_t \gets \textsc{Hill}(\mathbf{a}^{\mathrm{M}}_t)$ (Eq.~\ref{eq:hill_total}) \Comment{Hill dynamics}
        \State Step MuJoCo with $\boldsymbol{\tau}^{\text{musc}}_t$ \Comment{muscle-driven motion}
        \State Compute $r_t$  (Eq.~\ref{eq:reward}) \Comment{mimic reward}
        \State Append ($a_t, s_t, s_{t+1}, r_t$)  to $\mathbf{B}$
        \State Append $(\boldsymbol{\tau}^{\text{des}}_t, \boldsymbol{\tau}^{\text{musc}}_t)$ to $\mathbf{B}^\text{musc}$
    \EndFor

    \State \textbf{// Optimize three NNs}

    \State Update $\phi,\psi$ via PPO with $\mathbf{B}$
    \State Update $\theta$ via $L_{\mu}$ (Eq.~\ref{eq:muscle_loss}) and $\mathbf{B}^\text{musc}$
\EndFor
\State \Return $\pi_\phi$, $\mu_\theta$
\end{algorithmic}
\end{algorithm}

\subsubsection{Muscle Network}

The muscle network $\mu_\theta$ maps the desired joint torque $\boldsymbol{\tau}^{\text{des}}$ to per-muscle activations $\mathbf{a}^{\mathrm{M}}\!\in\![0,1]^{N_m}$, solving the muscle inverse problem. This map is \emph{redundant}: the $N_m$ muscles far outnumber the $N_\theta$ DOFs, so infinitely many activation patterns realize the same torque. Classical biomechanics resolves the redundancy per frame with a quadratic program (QP); we instead use a simple supervised network to solve it. Layer sizes and training hyperparameters are given in the supplementary material.

\textbf{Muscle-network loss.} The muscle network $\mu_\theta$ is trained online with three losses (Eq.~\ref{eq:muscle_loss}): a torque-realization term $L_{\text{torque}}$ that inverts the activation-to-torque map $\boldsymbol{\tau}^{\text{musc}}(\mathbf{a}^{\mathrm{M}}){=}\mathbf{L}\,\mathbf{a}^{\mathrm{M}}{+}\mathbf{b}$ (\S\ref{sec:physics}), a minimum-effort regularizer $L_{\text{reg}}$ that breaks the activation redundancy, and a physiological EMG prior $L_{\text{emg}}$. The first two are standard in muscle-actuated control; $L_{\text{emg}}$ is our contribution: existing muscle networks~\cite{lee_scalable_2019, feng_musclevae_2023, park_magnet_2025, li_towards_2026, myosuite} are trained on torque and effort alone, so their activations have no reason to match how a human recruits muscle (\S\ref{sec:res-emg}), whereas $L_{\text{emg}}$ steers the redundant solution toward a measured human EMG envelope. The three combine as
\begin{equation}
\begin{aligned}
  L_{\mu} = {}& \underbrace{\operatorname{mean}\!\Big[\big(\tfrac{1}{\kappa}(\boldsymbol{\tau}^{\text{musc}}(\mathbf{a}^{\mathrm{M}}) - \boldsymbol{\tau}^{\text{des}})\big)^2\Big]}_{L_{\text{torque}}} + w_{\text{reg}}\,\underbrace{\operatorname{mean}\big[(\mathbf{a}^{\mathrm{M}})^2\big]}_{L_{\text{reg}}} \\[0.5ex]
  &+ \lambda\,\underbrace{L_{\text{emg}}}_{\text{EMG prior}},
\end{aligned}
  \label{eq:muscle_loss}
\end{equation}
with torque scale $\kappa{=}100$, $w_{\text{reg}}{=}0.01$, and EMG weight $\lambda$ ($L_{\text{emg}}$ is defined in Eq.~\ref{eq:emg_prior}; setting $\lambda{=}0$ recovers the tracking-only loss for motions without a measured envelope). $L_{\text{reg}}$ is the $\sum a^2$ minimum-effort criterion of classical static optimization~\cite{anderson_static_2001}, which on its own drives the activations away from physiological EMG measurements.

\subsubsection{EMG Prior}
\label{sec:emg-prior}

\begin{figure}[htbp]
    \centering
    \includegraphics[width=1.0\linewidth]{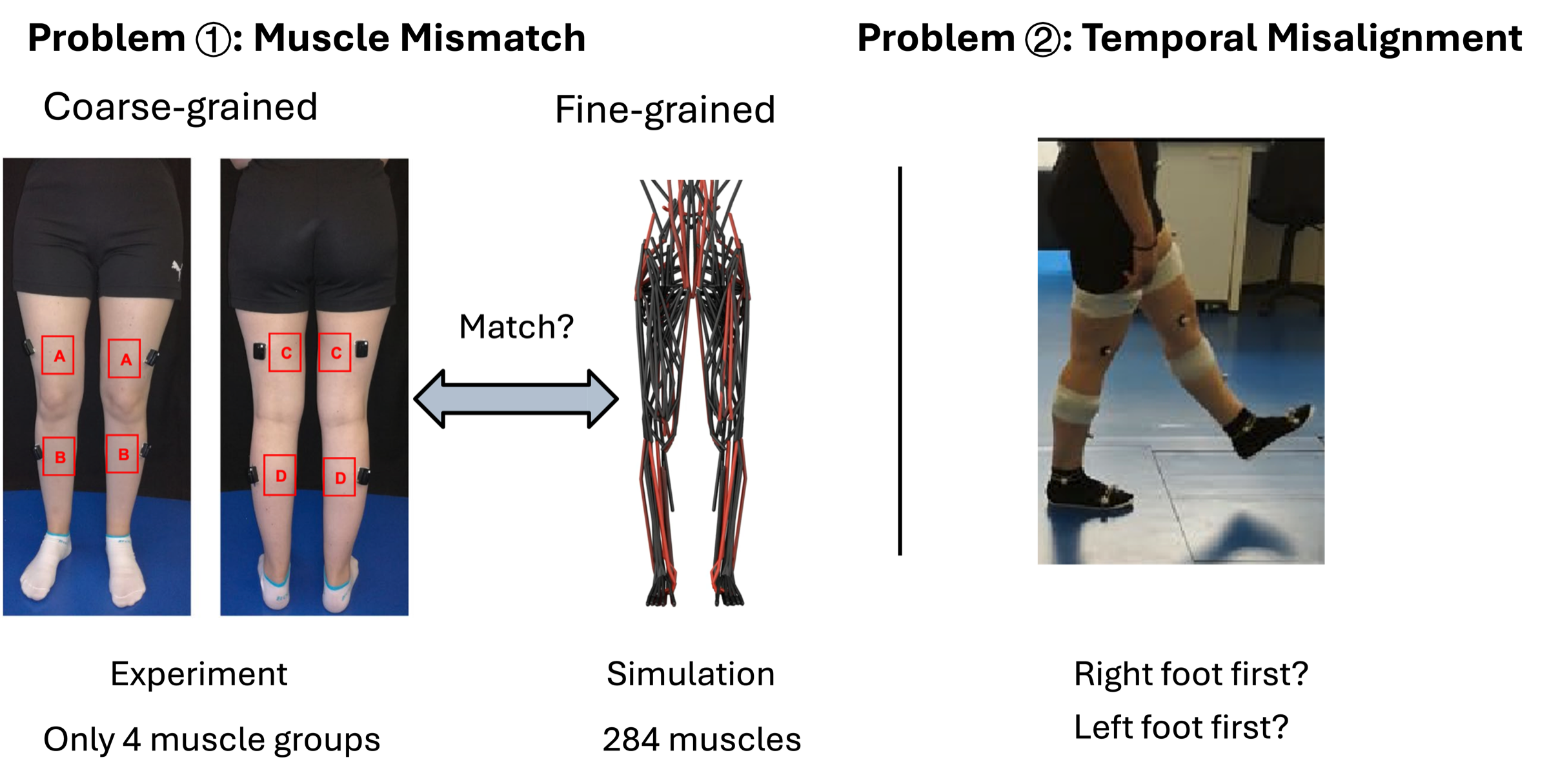}
    \caption{Challenges in aligning the experimental data~\cite{moreira2021lower} with our simulation: 1)~Muscle mismatch: the measured data is coarse-grained, with only four right-leg electrodes, each standing for a synergist group (anterior view, A:~vastus lateralis, quadriceps; B:~tibialis anterior, dorsiflexors; posterior view, C:~biceps femoris, hamstrings; D:~lateral gastrocnemius, triceps surae), while our computational model is fine-grained, carrying one activation per muscle. 2)~Temporal misalignment: the measured data is not aligned with the reference motion, e.g.\ the alignment depends on which foot touches the ground first.}
    \label{fig:emg-prior}
\end{figure}

So far, we have obtained activations that generate physically plausible motion through muscle forces, but these activations are still physiologically implausible. We therefore steer the redundant solution toward real surface electromyography (sEMG) from Moreira et al.~\cite{moreira2021lower}. Two mismatches stand between that recording and our model (Fig.~\ref{fig:emg-prior}). The first is a \emph{muscle mismatch}: the recording is coarse-grained, using only four right-leg electrodes, whereas our model is fine-grained. The second is a \emph{temporal misalignment}: the recording is not aligned with the reference motion, so the two cannot be compared sample by sample. We apply two treatments, detailed in turn below: 1)~we match each electrode not to a single fiber but to a whole synergist group; 2)~we fold both signals onto a common gait cycle and compare only their shape and timing, not their magnitude.

\textbf{Synergist-group mean.} Let $\mathcal{C}$ be the EMG channels (the four right-leg surface electrodes of Moreira et al.~\cite{moreira2021lower}). A surface electrode integrates over a muscle region and the redundant solver distributes load within a synergist group, so channel $c$ is matched not to one fiber but to the cross-section-weighted mean activation of its synergist group $G_c$:
\begin{equation}
  \bar a_c = \sum_{m\in G_c} \omega_{c,m}\, a^{\mathrm{M}}_m,
  \qquad
  \omega_{c,m} = \frac{f_{0,m}}{\sum_{m'\in G_c} f_{0,m'}},
  \label{eq:group_act}
\end{equation}
where $f_{0,m}{=}\sigma_0 A^{\mathrm{M}}_m$ is the maximal isometric force of muscle $m$ (Eq.~\ref{eq:musc_force}), obtained from its physiological cross-sectional area $A^{\mathrm{M}}_m$ and the maximal isometric stress $\sigma_0$, the peak stress a fully activated fiber can sustain. Since $\sigma_0$ is uniform across muscles (value in the supplementary), the weight reduces to a ratio of cross-sectional areas, so the stronger muscles of a group dominate its mean. Matching the group mean rather than a single fiber also guards against a dead-unit pathology (supplementary).

\textbf{Gait-cycle alignment.} Each rollout sample carries a gait phase $\phi = \operatorname{fmod}(t,\,T)/T \in [0,1)$ recovered from its simulation time $t$, with $T$ the gait-cycle period, and is placed on the $0\!-\!100\%$ cycle as $\big((\phi - \phi_{\text{hs}})\bmod 1\big)\cdot 100$ to align it to right heel strike~\cite{zeni2008two}, where $\phi_{\text{hs}}{=}0.254$ is the heel-strike offset calibrated by forward kinematics on the reference foot. The prior is a phase-indexed MSE that pulls each group activation toward the peak-normalized human envelope $\hat e_c(\phi)$:
\begin{equation}
  L_{\text{emg}} = \frac{1}{|\mathcal{C}|}\sum_{c\in\mathcal{C}} w_{\text{emg},c}\,\big(\bar a_c - s\,\hat e_c(\phi)\big)^2,
  \label{eq:emg_prior}
\end{equation}
where $s$ scales the unit-peak envelope into the activation range and $w_{\text{emg},c}$ weights each channel (the weight $\lambda$ against torque realization is set in Eq.~\ref{eq:muscle_loss}). The prior enters \emph{only} the muscle-network loss; the policy's motion objective (Eq.~\ref{eq:reward}) is untouched, so it re-selects \emph{which} activation realizes a motion without changing the motion itself. We use $\lambda{=}1$, $s{=}0.3$, and $w_{\text{emg},c}{=}1$ for all channels; $\hat e_c$ is the $13$-subject Moreira cohort mean on a $101$-bin gait grid. It is enabled wherever a measured envelope exists (the walk) and switched off ($\lambda{=}0$) otherwise; its effect on physiological plausibility is evaluated in \S\ref{sec:res-emg}.

\subsection{1D Hill-type Muscle Surrogate}
\label{sec:physics}

\begin{figure}[htbp]
    \centering
    \includegraphics[width=1.0\linewidth]{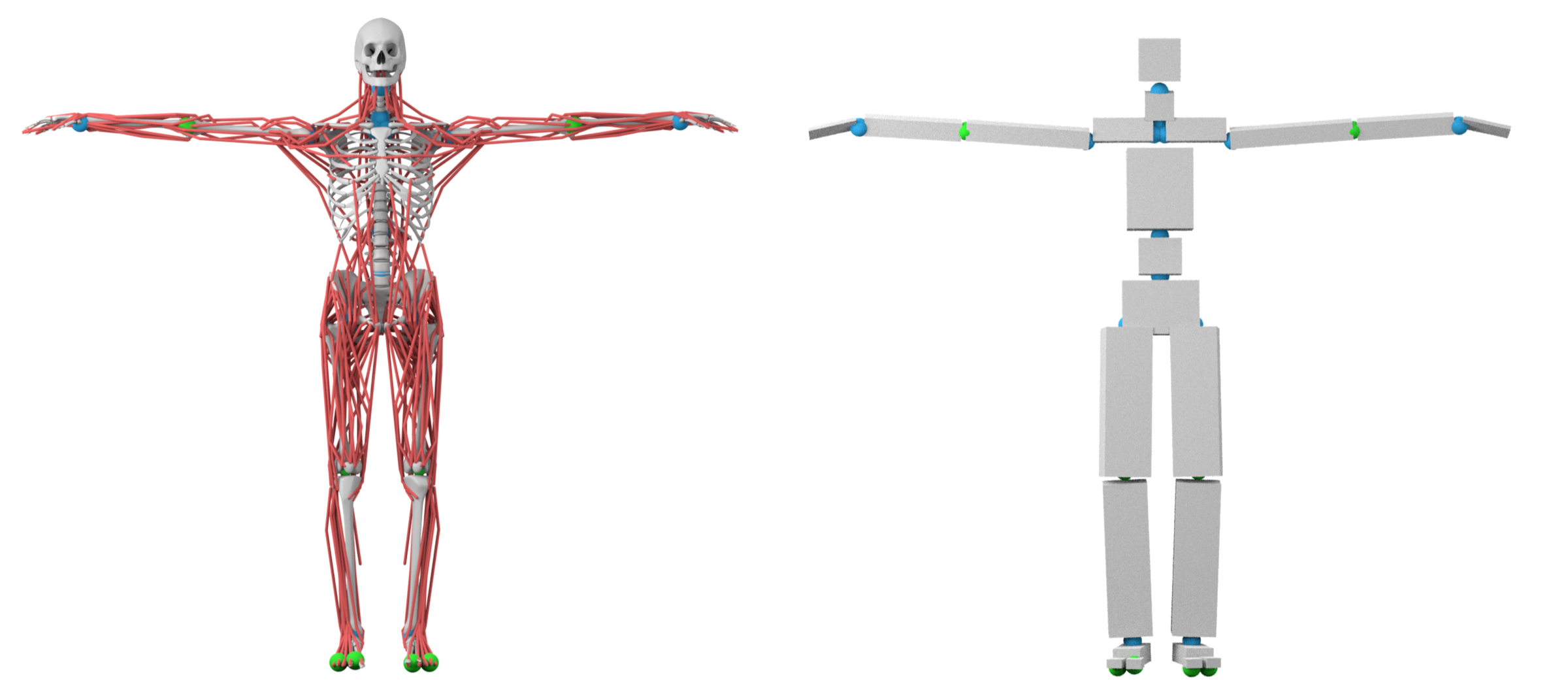}
    \caption{The 1D musculoskeletal asset at rest. \emph{Left:} the rigid skeleton with $284$ 1D Hill-type muscles (red polylines) and 22 joints modeled as spherical (blue) or revolute (green). \emph{Right:} the corresponding MuJoCo rigid bodies carrying mass and inertia, with the 1D muscle anchor points.}
    \label{fig:1d-asset}
\end{figure}

For motion to be truly actuated by muscle forces, the musculoskeletal system should closely match human anatomy. However, we empirically find that 3D volumetric muscles are overly flexible during RL training due to their excessive number of degrees of freedom, leading to unstable learning. Lee et al.~\cite{lee_dexterous_2018} run the volumetric solver directly inside their optimization loop, which was also our initial idea. However, in our experiments, this approach was unstable and failed to converge. Therefore, we instead train against a 1D Hill-type muscle model~\cite{lee_scalable_2019, feng_musclevae_2023, park_magnet_2025, li_towards_2026} as a surrogate, and then lift it to a 3D volumetric muscle model in the next section.  The Hill-type model plays a dual role in our system: (i)~as the \emph{1D musculotendon surrogate} that drives RL training, its scalar tension routed through MuJoCo tendon paths to the skeleton (Alg.~\ref{alg:train}); and (ii)~as the \emph{constitutive law} for the 3D XPBD volumetric muscle of \S\ref{sec:xpbd}, injected there as an energy-based fiber constraint (Eq.~\ref{eq:fiber_constraint}). We give the shared 1D formulation here and the 3D lift in \S\ref{sec:xpbd}. The 1D musculoskeletal asset (Fig.~\ref{fig:1d-asset}) is a rigid-body skeleton with spherical and revolute joints, actuated by 1D Hill-type muscles routed as polyline tendon paths between bone attachment sites.

\begin{figure}[t]
    \centering
    \includegraphics[width=\linewidth]{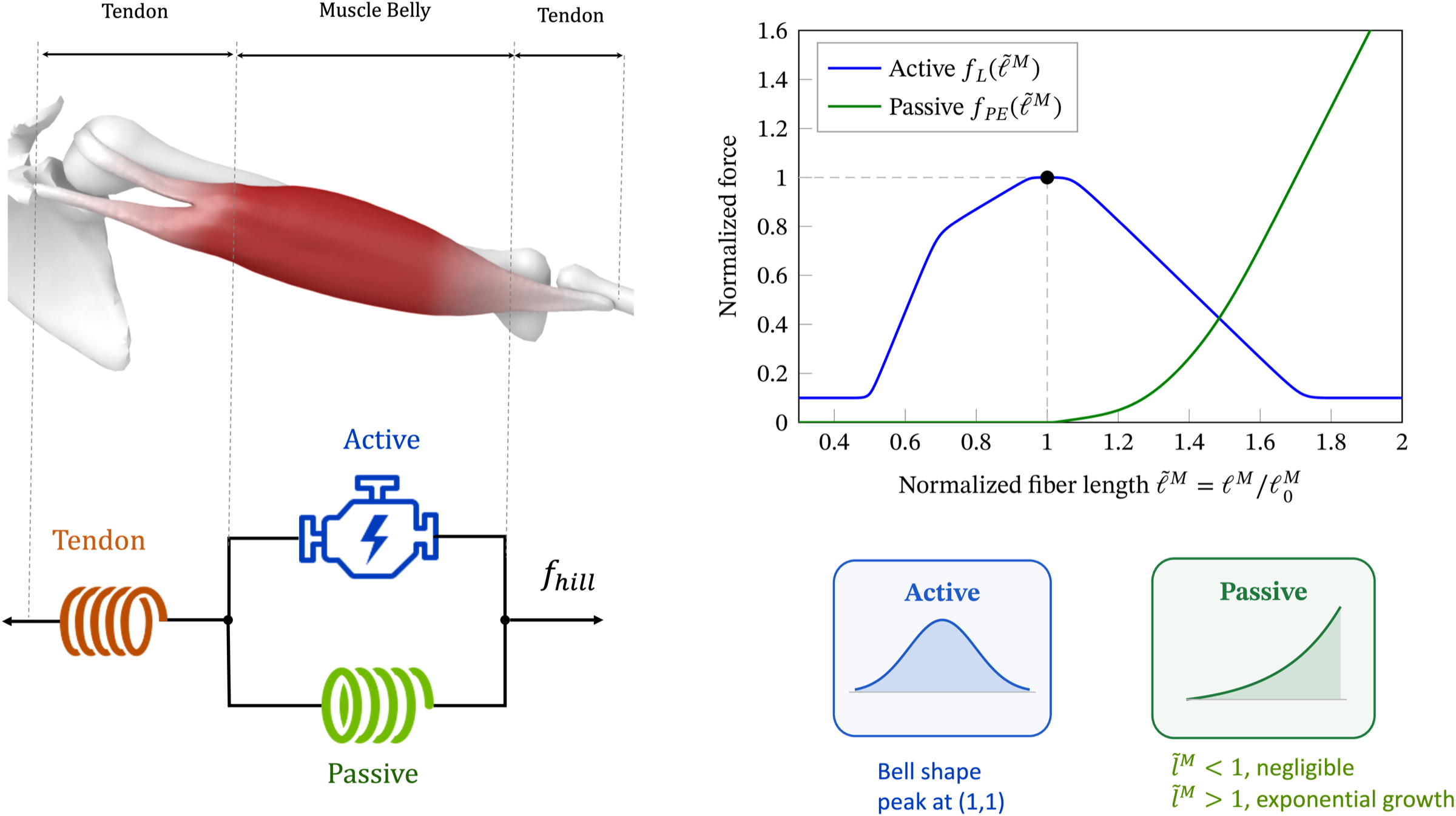}
    \caption{Schematic of the Hill-type musculotendon model: a contractile (active) element parallel to an elastic (passive) element, in series with an elastic tendon. The active force--length curve peaks at the optimal fiber length ($\tilde{\ell}^{\mathrm{M}}{=}1$) and is scaled by activation.}
    \label{fig:Hill-model}
\end{figure}

The Hill-type model (Fig.~\ref{fig:Hill-model}) places a tendon spring in series with a muscle belly, itself an active contractile element parallel to a passive spring. Each fiber generates a normalized scalar force:
\begin{equation}
  f_{\text{hill}} =
    \underbrace{\textcolor{hillActive}{a \, f^{\mathrm{L}}(\tilde{\ell}^{\mathrm{M}})\, f^{\mathrm{V}}(\dot{\ell}^{\mathrm{M}})}}_{\textcolor{hillActive}{\text{active}}}
    + \underbrace{\textcolor{hillPassive}{f^{\mathrm{PE}}(\tilde{\ell}^{\mathrm{M}})}}_{\textcolor{hillPassive}{\text{passive}}}
    + \underbrace{\textcolor{hillTendon}{f^{\mathrm{T}}(\tilde{\ell}^{\mathrm{T}})}}_{\textcolor{hillTendon}{\text{tendon}}},
  \label{eq:hill_total}
\end{equation}
where $a \in [0,1]$ is the muscle activation and $\tilde{\ell}^{\mathrm{M}}$ the fiber length normalized by the optimal fiber length, so $f_{\text{hill}}$ is a dimensionless multiplier equal to $1$ at that length (the 3D counterpart of $\tilde{\ell}^{\mathrm{M}}$, the along-fiber stretch $\lambda_{\mathrm{f}}$, is defined in \S\ref{sec:xpbd}).  We take the force--length curves $f^{\mathrm{L}}$ and $f^{\mathrm{PE}}$ from Millard et al.~\cite{millard_flexing_2013} and evaluate them through a look-up table to accelerate RL training; their piecewise-B\'{e}zier form and parameters are given in the supplementary. 
For simplicity we further assume 1)~an infinitely stiff tendon and 2)~no velocity-dependent force $f^{\mathrm{V}}$, since only conservative forces admit an energy potential (\S\ref{sec:fiber-constraint}). We obtain the 1D muscle force by multiplying $f_{\text{hill}}$ by the peak isometric force $f_0$, which depends on the muscle (e.g., leg muscles are stronger than arm muscles). $f_0 = \sigma_0 A^{\mathrm{M}}$ (physiological cross-sectional area $A^{\mathrm{M}}$). The resulting 1D Hill-type muscle force is
\begin{equation}
  F(a) = f_0 \big( a\, f^{\mathrm{L}}(\tilde{\ell}^{\mathrm{M}}) + f^{\mathrm{PE}}(\tilde{\ell}^{\mathrm{M}}) \big).
  \label{eq:musc_force}
\end{equation}

\textbf{Muscle force to joint torque.} Summing the segments of the anchor polyline gives the path length, hence the normalized fiber length $\tilde{\ell}^{\mathrm{M}}$ that Eq.~\ref{eq:hill_total} evaluates; mapping the anchor forces through their point Jacobians makes the total torque affine in activation, $\boldsymbol{\tau}^{\text{musc}}(\mathbf{a}^{\mathrm{M}}) = \mathbf{L}\,\mathbf{a}^{\mathrm{M}} + \mathbf{b}$, with $\mathbf{L}$ the per-unit-activation moment arms and $\mathbf{b}$ the passive torque (root DOFs excluded; both derived in the supplementary). This closed-form, differentiable map is the Hill forward through which $\mu_\theta$ is trained (\S\ref{sec:rl}) and the torque that drives MuJoCo.

\subsection{3D Volumetric Muscle Simulation}
\label{sec:xpbd}

To achieve a stable, controllable, and realistic volumetric muscle simulation, three key challenges must be addressed: 1)~anisotropic fiber force, 2)~volume conservation, and 3)~muscle--bone coupling. The first two drive the realism: the fiber force contracts the muscle along the fibers, while volume conservation makes it bulge in the orthogonal directions. Enforcing volume conservation is in turn limited by solver convergence: under the high stiffness muscle requires, an under-converged solve leaves visible artifacts and inverted tetrahedra. Muscle--bone coupling is governed by attachments: attachments that are too stiff distort the muscle, while those that are too soft fail to transmit force. We address them in turn. We embed the Hill-type curve into the constitutive law, so that per-fiber activation directly shortens the muscle along its fibers. We propose a GMG (Geometric Multigrid) solver for fast and stable simulation (\S\ref{sec:gmg}). Finally, we partition each muscle by masks into tendon and belly regions, attaching the tendons hard and the belly soft. We simulate the volumetric muscle with XPBD~\cite{macklin_xpbd_2016}, giving first the solver update and the geometric multigrid that accelerates it, then the individual constraints.

\subsubsection{XPBD Update}
\label{sec:xpbd-update}
The volumetric muscle deformation is driven by a set of position-based constraints: volume and shape preservation (\S\ref{sec:vol-shape}), anisotropic fiber contraction (\S\ref{sec:fiber-constraint}), attachment (\S\ref{sec:attach}), and collision (\S\ref{sec:collision}). Each is solved with the standard XPBD update:
\begin{align}
  \Delta \lambda &= \frac{-C - \tilde{\alpha} \, \lambda}{\sum_i w_i \lVert \nabla C \rVert^2 + \tilde{\alpha}}, \\
  \Delta \mathbf{x}_i &= w_i \, \nabla C \, \Delta \lambda.
  \label{eq:xpbd_update}
\end{align}
Here $C$ is one of the constraint functions and $\nabla C$ its gradient with respect to the participating vertex positions. $\lambda$ is the constraint's accumulated Lagrange multiplier (reset to zero at the start of each time step) and $\Delta\lambda$ is its per-iteration increment. $\tilde{\alpha} = 1/(k\,\Delta t^2)$ is the compliance, i.e.\ the inverse of the constraint stiffness $k$ scaled by the squared time step $\Delta t$; a stiffer constraint thus has a smaller $\tilde{\alpha}$. $w_i$ is the inverse mass of vertex $i$, and $\Delta\mathbf{x}_i$ is the position correction applied to that vertex.  Once all constraints have been projected, velocities are recovered from the position change over the step, $\mathbf{v}^f \gets (\mathbf{x}^f - \mathbf{x}^f_{\text{prev}})/\Delta t$. The individual constraints are defined next; each constraint's gradient is derived in the supplementary.

\subsubsection{Geometric Multigrid}
\label{sec:gmg}
\begin{figure}[htbp]
    \centering
    \includegraphics[width=\linewidth]{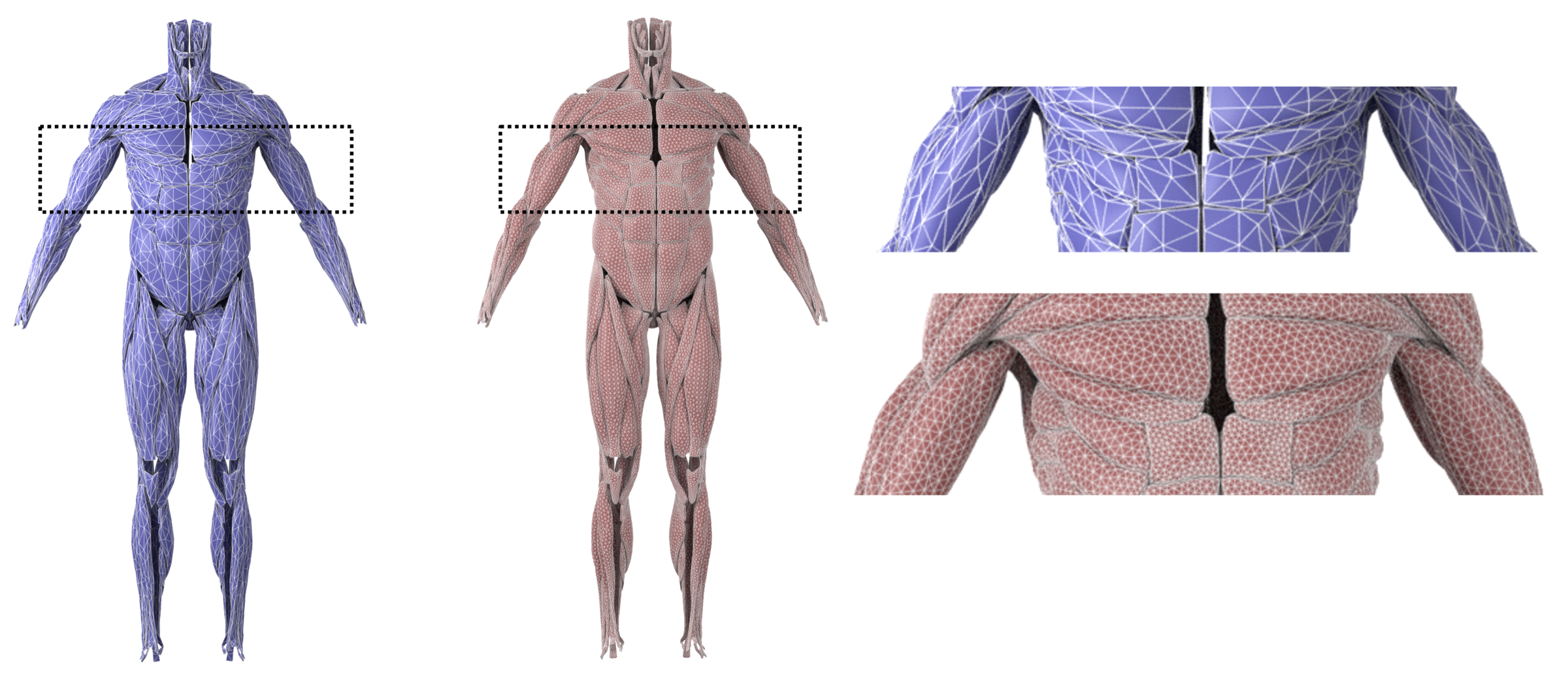}
    \caption{The 3D volumetric muscle asset: the coarse ``cage'' mesh ($31{,}044$ tetrahedra, left) and the fine simulation mesh ($468{,}162$ tetrahedra, right).}
    \label{fig:coarseFine}
\end{figure}

The geometric multigrid is a core contribution of this work: it is what makes our stiff, high-resolution volumetric muscle solve both stable and fast. Previous position-based hierarchies, such as HPBD~\cite{Muller2008-HPBD}, have two limitations: 1)~they are formulated for cloth rather than tetrahedral meshes; 2)~they do not close a complete V-cycle, so they act more like a low-resolution driver for the high-resolution mesh. Therefore, we develop a new algorithm for geometric multigrid XPBD that supports tetrahedral meshes with a Petrov--Galerkin restriction. We construct the two-level hierarchy by geometric coarsening: we first inflate the fine surface triangular mesh slightly, remesh it to a coarser triangle mesh, and then tetrahedralize it (Fig.~\ref{fig:coarseFine}). Our core insight is that the coarse tetrahedron acts as a ``cage'' that controls the motion of the fine vertices. The prolongator $\mathbf{P}$ is established by barycentric coordinates,
\begin{align}
    \mathbf{x}^f &= \mathbf{x}_0^f + \sum_{i=1}^4 w_i\, \Delta\mathbf{x}^c_i, \label{eq:prolongation}
\end{align}
where $w_1,\dots,w_4$ are the barycentric coordinates of the fine vertex with respect to the coarse tetrahedron, with $w_4=1-\sum_{i=1}^3 w_i$. $\mathbf{x}_0^f$ is the fine position before the multigrid update. Each set of four weights for a tetrahedron defines a row of $\mathbf{P}$.  As in HPBD~\cite{Muller2008-HPBD}, we prolongate the coarse correction $\Delta \mathbf{x}^c$ instead of the direct coarse solution $\mathbf{x}^c$.

We define restriction as the least-squares state that best reproduces the fine state $\mathbf{x}^f$ within the cage subspace,
\begin{equation}
    \mathbf{x}^c=\operatorname*{arg\,min}_{\mathbf{x}^c}\;\bigl\lVert \mathbf{P}\,\mathbf{x}^c-\mathbf{x}^f \bigr\rVert_2^2,
\end{equation}
whose normal equations give the Petrov--Galerkin restriction
\begin{equation}
    \mathbf{x}^c=\underbrace{\bigl(\mathbf{P}^{\!\top}\mathbf{P}\bigr)^{-1}\mathbf{P}^{\!\top}}_{\textstyle \mathbf{R}}\,\mathbf{x}^f.
    \label{eq:restriction}
\end{equation}
Since $\mathbf{P}$ has full column rank, $\mathbf{R}$ is the \emph{left pseudo-inverse} of $\mathbf{P}$, so restriction exactly recovers any coarse-level vector after prolongation, i.e., $\mathbf{R}\mathbf{P}=\mathbf{I}$. A desirable property of this Petrov--Galerkin restriction is that it has $\mathbf{R}\mathbf{P}=\mathbf{I}$, which is why we choose it over the Galerkin restriction $\mathbf{R}=\mathbf{P}^{\!\top}$.

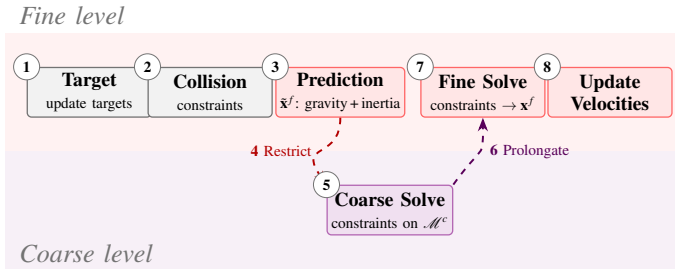
\begin{figure}[htbp]
\centering
\resizebox{\columnwidth}{!}{%
\begin{tikzpicture}[
    font=\small,
    >={Stealth[length=2.2mm]},
    box/.style={
        rounded corners=2pt, draw=black!70, line width=0.6pt,
        minimum height=8mm, minimum width=20mm, align=center,
        inner sep=2pt, fill=white},
    fine/.style={box, fill=red!10, draw=red!60},
    coarse/.style={box, fill=violet!12, draw=violet!60},
    setup/.style={box, fill=black!5, draw=black!55},
    flow/.style={->, line width=0.8pt, black!75},
    down/.style={->, line width=0.9pt, red!70!black, dashed},
    up/.style={->, line width=0.9pt, violet!70!black, dashed},
    lane/.style={font=\large\itshape, text=black!55},
    num/.style={circle, draw=black!55, fill=white, inner sep=0pt,
        minimum size=4.2mm, font=\scriptsize\bfseries}
]
\begin{scope}[on background layer]
    \fill[red!5]    (-0.8,0.95) rectangle (10.0,-0.95);
    \fill[violet!6] (-0.8,-0.95) rectangle (10.0,-2.85);
\end{scope}
\node[setup] (bone)  at (0.55,0) {\textbf{Target}\\\scriptsize update targets};
\node[setup] (coll)  at (2.5,0)  {\textbf{Collision}\\\scriptsize constraints};
\node[fine]  (pred)  at (4.6,0)  {\textbf{Prediction}\\\scriptsize $\tilde{\mathbf{x}}^f$: gravity\,+\,inertia};
\node[fine]  (fsolve) at (6.9,0) {\textbf{Fine Solve}\\\scriptsize constraints $\to \mathbf{x}^f$};
\node[fine]  (vel)   at (8.95,0) {\textbf{Update}\\\textbf{Velocities}};
\node[coarse] (csolve) at (5.4,-1.9) {\textbf{Coarse Solve}\\\scriptsize constraints on $\mathcal{M}^c$};
\node[lane, anchor=west] at (-0.7,1.25) {Fine level};
\node[lane, anchor=west] at (-0.7,-2.6) {Coarse level};
\draw[down] (pred.south) .. controls +(0,-0.7) and +(-0.5,0.7) .. (csolve.north west)
    node[pos=0.5, left=1pt, text=red!70!black, font=\scriptsize] {\textbf{4} Restrict};
\draw[up] (csolve.north east) .. controls +(0.5,0.7) and +(0,-0.7) .. (fsolve.south)
    node[pos=0.5, right=1pt, text=violet!70!black, font=\scriptsize] {\textbf{6} Prolongate};
\node[num] at (bone.north west)   {1};
\node[num] at (coll.north west)   {2};
\node[num] at (pred.north west)   {3};
\node[num] at (csolve.north west) {5};
\node[num] at (fsolve.north west) {7};
\node[num] at (vel.north west)    {8};
\end{tikzpicture}%
}
\caption{The per-step volumetric muscle simulation as a two-level geometric-multigrid V-cycle. Steps \textbf{4}~restriction, \textbf{5}~coarse solve and \textbf{6}~prolongation use Eq.~\eqref{eq:restriction}, Eq.~\eqref{eq:xpbd_update} and Eq.~\eqref{eq:prolongation} respectively.}
\label{fig:musclesim-vcycle}
\end{figure}

Our restriction and prolongation processes are matrix-free and do not require explicit construction of the large sparse matrices $\mathbf{P}$ and $\mathbf{R}$. Because every fine vertex is embedded in the cage of a single muscle, $\mathbf{P}$, and hence $\mathbf{P}^{\!\top}\mathbf{P}$, is block-diagonal across muscles, so Eq.~\eqref{eq:restriction} decouples into one small dense solve per muscle. We therefore precompute and store the per-muscle pseudo-inverse blocks $(\mathbf{P}^{\!\top}\mathbf{P})^{-1}\mathbf{P}^{\!\top}$ once, reducing restriction to a block-diagonal matrix--vector product at run time. The per-step volumetric simulation is summarized in Figure~\ref{fig:musclesim-vcycle}. Note that the multigrid can be easily switched off by skipping the restriction, coarse-solving, and prolongation steps, which reduces it to the classic XPBD algorithm.

\subsubsection{Volume and Shape Preservation Constraints}
\label{sec:vol-shape}
Incompressibility is enforced \emph{primarily} by a Stable Neo-Hookean (SNH) material~\cite{smith_stable_2018}, cast as a coupled pair of XPBD constraints following Macklin and M\"{u}ller~\cite{macklin_snh_2021}, alongside an explicit volume constraint and an ARAP shape term:
\begin{align}
   \text{SNH Deviatoric}:\; C_{\text{dev}} &= \lVert\mathbf{F}\rVert_F - \sqrt{3}, \\
   \text{SNH Hydrostatic}:\; C_{\text{hyd}} &= \det\mathbf{F} - \alpha, \quad \alpha = 1 + \mu/\lambda, \\
   \text{Volume}:\; C_{\text{vol}} &= V - V_0, \\
   \text{ARAP}:\; C_{\text{arap}} &= \lVert \mathbf{F} - \mathbf{R} \rVert_F,
\end{align}
where $\mathbf{F}$ is the deformation gradient and $\mathbf{R}$ its rotation part, $V$ and $V_0$ the current and rest volume, $\mu$ and $\lambda$ the shear and bulk moduli, and $\alpha$ a rest-stability offset that makes the undeformed configuration an exact rest state of the energy. In SNH the hydrostatic term enforces incompressibility while the deviatoric term carries the shear response; the ARAP (As-Rigid-As-Possible) term further resists shear and preserves shape. For that term we found an SVD-based formulation with non-negative clamping~\cite{irving_invertible_2004} more robust than the polar-decomposition one.

\subsubsection{Anisotropic Fiber Constraints}
\label{sec:fiber-constraint}

We inject the 1D Hill law into the 3D medium as an energy-based fiber constraint. Whereas Lee et al.~\cite{lee_dexterous_2018} realize the active fiber force through a first Piola--Kirchhoff stress, we formulate it at the position level for XPBD. The normalized fiber length of Eq.~\ref{eq:hill_total} is lifted from the 1D path measurement (\S\ref{sec:physics}) to the stretch of the deformation gradient $\mathbf{F}$ along the muscle's rest fiber direction:
\begin{equation}
  \lambda_{\mathrm{f}} = \lVert \mathbf{F}\,\mathbf{d} \rVert \;\equiv\; \tilde{\ell}^{\mathrm{M}},
  \label{eq:fiber_stretch}
\end{equation}
where $\mathbf{d}$ is the unit fiber direction in the rest configuration. This 3D stretch is \emph{identically} the 1D normalized fiber length: both are the current fiber length divided by its optimal (rest) length, so we keep $\tilde{\ell}^{\mathrm{M}}$ for the 1D Hill curves and write $\lambda_{\mathrm{f}}$ for the same quantity in the 3D continuum. XPBD stores each elastic potential as the quadratic form $\Psi=\tfrac{1}{2}k\,C^2$ with compliance $\tilde{\alpha}=1/(k\,V_0\,\Delta t^2)$, so setting $\tfrac{1}{2}C_{\text{fiber}}^2=\Psi_{\text{fiber}}$ encodes the Hill strain energy directly in the constraint:
\begin{equation}
  \Psi_{\text{fiber}} = \sigma_0 \!\!\int_{\lambda_{\mathrm{f},\min}}^{\lambda_{\mathrm{f}}}\!\! f_{\text{hill}}(s) \, ds,
  \qquad
  C_{\text{fiber}} = \sqrt{2\,\Psi_{\text{fiber}}},
  \label{eq:fiber_constraint}
\end{equation}
where $f_{\text{hill}}$ is given by Eq.~\ref{eq:hill_total}, from which we drop $f^{\mathrm{V}}$, $f^{\mathrm{T}}$, and $f^{\mathrm{PE}}$ in the 3D medium. Because $f_{\text{hill}}$ is a piecewise quintic polynomial, the energy integral has a \emph{closed-form} degree-10 polynomial per segment, avoiding numerical quadrature. Injecting the \emph{same} Hill law that trains the policy makes the volumetric contraction obey exactly the constitutive relation the activations were solved against, and because only the along-fiber component is constrained, activation shortens the belly along $\mathbf{d}$ while volume conservation drives the orthogonal bulge. An anisotropic ARAP alternative is compared in \S\ref{sec:abl-fiber}; both constraints, their gradients, and the fiber stiffness are given in the supplementary.

\subsubsection{Attachment Constraints}
\label{sec:attach}
The muscle is coupled to the skeleton by attachment constraints that drive attachment vertices toward their targets on the bone:
\begin{equation}
    C_{\text{attach}} = \lVert \mathbf{x}_i - \mathbf{x}_i^* \rVert - L_0,
\end{equation}
where $\mathbf{x}_i$ is the position of an attached vertex, $\mathbf{x}_i^*$ its target on the bone, and $L_0$ the rest length between the vertex and its target. The bone--muscle coupling is one-way, the rigid bone driving the deformable muscle, with targets found by a nearest-point search on the bone surface at the rest pose. Neighboring muscles are additionally tied to one another by inter-muscle springs, built once at the rest pose between nearby surface vertices of adjacent muscles; these keep the musculature bundled and are two-way, unlike the bone attachments. We use two kinds of attachment: \emph{pins}, which are hard and impose the position directly, and \emph{springs}, which are soft and allow some deformation. A user-painted mask assigns pins to the tendon regions near the origin and insertion and springs to the belly. The two are respectively Dirichlet and Robin boundary conditions.

Their balance is a trade-off: too many pins drive up tetrahedron inversions and leave the surface unsmooth, while too few let the muscle lag behind and detach from the bone, most visibly on the thin, slender muscles that are dominated by these boundary conditions rather than by elasticity (\S\ref{sec:gmg}).

\subsubsection{Collision Handling}
\label{sec:collision}

We handle muscle--bone collisions with an SDF~\cite{macklin_local_2020}. Each bone is rigid, so its SDF $\phi_b$ is precomputed once at the rest pose and queried in the bone's local frame, needing no per-substep tessellation or nearest-primitive search. A muscle surface vertex $\mathbf{x}_p$ found to penetrate ($\phi_b<0$) is resolved by a \emph{unilateral} constraint projected alongside the other XPBD constraints,
\begin{equation}
  C_{\text{col}} = \phi_b(\mathbf{x}_p), \qquad
  \nabla C_{\text{col}} = \nabla \phi_b(\mathbf{x}_p),
  \label{eq:collision}
\end{equation}
which pushes the vertex back onto the bone surface along the SDF gradient and is released once it leaves the bone. We omit explicit muscle--muscle collision: neighboring muscles are already held apart by inter-muscle springs, and since the musculature is primarily bone-driven the residual inter-muscle penetration is minor. The broad/narrow-phase details and this trade-off are given in the supplementary.

\subsection{Motion Retargeting and Muscle Mapping}
\label{sec:retarget}

The RL stage operates on a 1D musculoskeletal asset, whereas the volumetric simulation runs on a 3D character with its own skeletal proportions and tetrahedral muscle meshes. Two transfer steps bridge the two representations (Fig.~\ref{fig:pipeline}). \textbf{Motion Retargeting}: The 1D skeleton spans $N_\theta{=}50$ actuated DOFs over $22$ joints, the floating root aside ($14$ spherical joints at $3$~DOF and $8$ revolute joints at $1$~DOF, i.e.\ $14{\times}3+8{\times}1=50$). We first perform a joint-level retargeting to transfer the 1D skeleton motion onto the 3D skeleton, where 20 key points are selected and then full-body inverse kinematics is solved to recover the 3D joint angles. \textbf{Muscle Mapping}: Usually the 3D model and the 1D model do not have a one-to-one correspondence, so we similarly need to map the $284$ 1D Hill muscles onto the $194$ 3D volumetric muscles. The two assets differ in muscle count because they originate from different musculoskeletal models. We map them by anatomical semantic similarity and spatial correspondence, i.e., matching by anatomical name first and, for any muscle left unmatched, falling back to the nearest one. This step is optional if the two assets are already in one-to-one correspondence.

\section{Results}
\label{sec:results}

We evaluate our system in four parts: a full-body demonstration on four motions (\S\ref{sec:res-motion}), validation of the discovered activations against human sEMG (\S\ref{sec:res-emg}), comparisons against production and neural volumetric muscle systems (\S\ref{sec:res-comparison}), and an ablation of each component (\S\ref{sec:res-ablation}). All RL policies are trained on the 1D Hill surrogate and the 3D volumetric simulation is run as a single forward pass at inference (\S\ref{sec:pipeline}).

RL co-training is implemented in PyTorch, and the articulated body simulation used for training in MuJoCo~\cite{mujoco}, with 64 parallel environments per motion; the 3D simulation is implemented in Warp~\cite{nvidia_warp_2023}, is fully GPU accelerated, and runs on a single NVIDIA RTX~4090 with an AMD Ryzen 9 5950X CPU. We adopt USD (Universal Scene Description) as the interchange format for all assets (skeletons, muscle tetrahedral meshes, fibers, and simulation caches), and render the final scenes in Houdini~\cite{Houdini}. Training and simulation parameters are given in the supplementary.

\subsection{Full-Body Demonstration}
\label{sec:res-motion}

\begin{figure*}[htbp]
    \centering
    \subfloat[Walk\label{fig:motion-walk}]{\includegraphics[width=0.2376\textwidth]{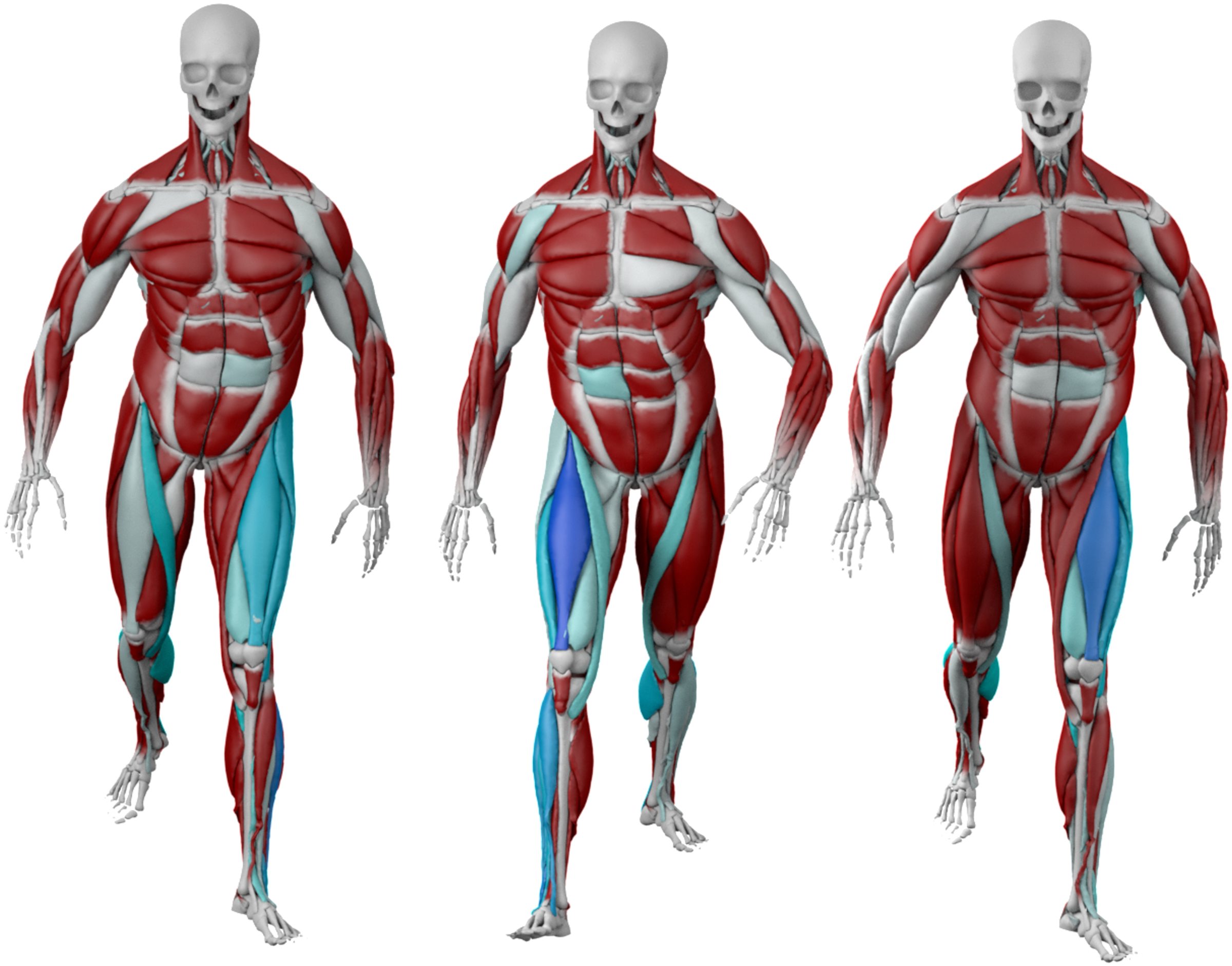}}\hfill
    \subfloat[Kick\label{fig:motion-kick}]{\includegraphics[width=0.2691\textwidth]{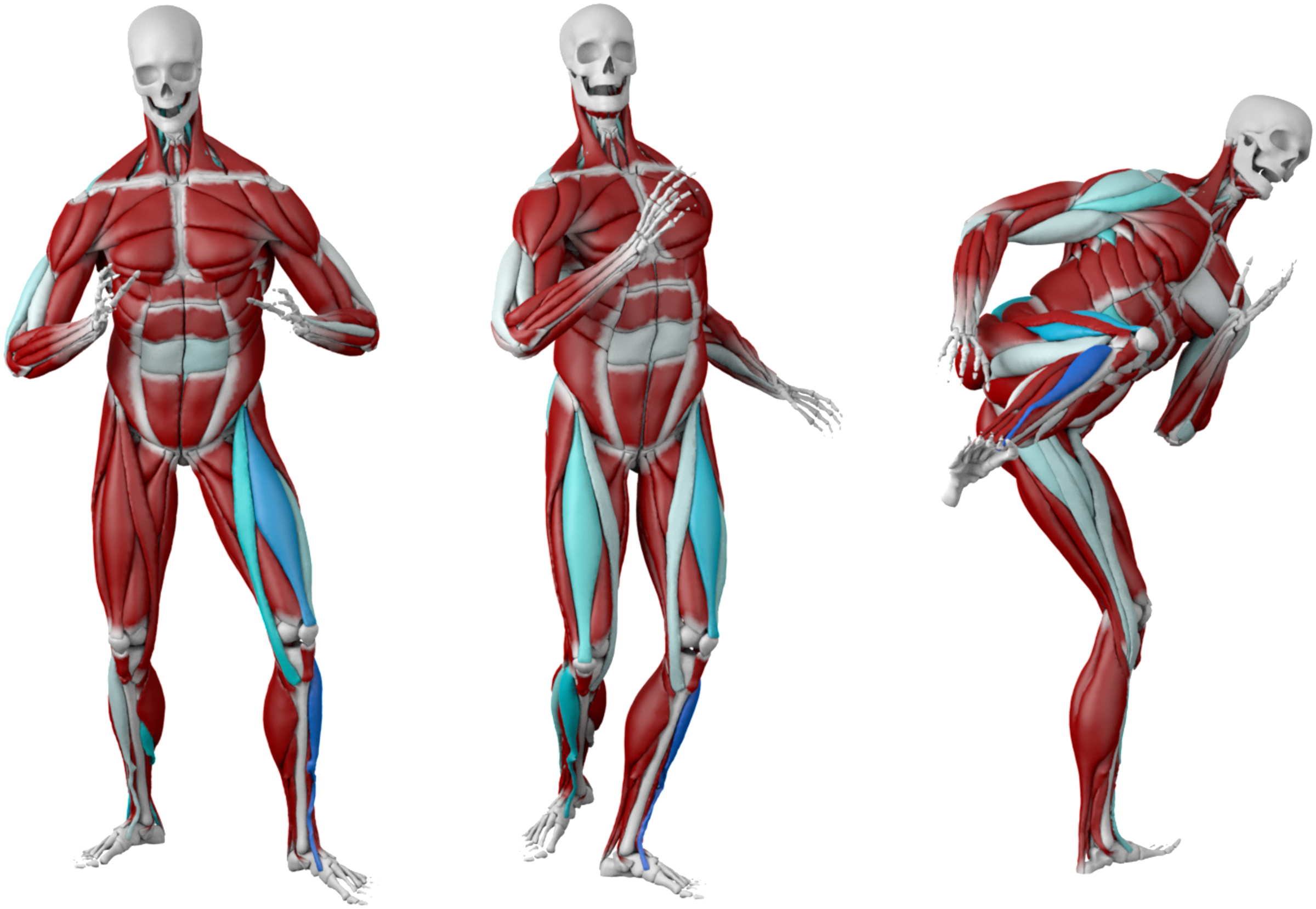}}\hfill
    \subfloat[Swim\label{fig:motion-swim}]{\includegraphics[width=0.2153\textwidth]{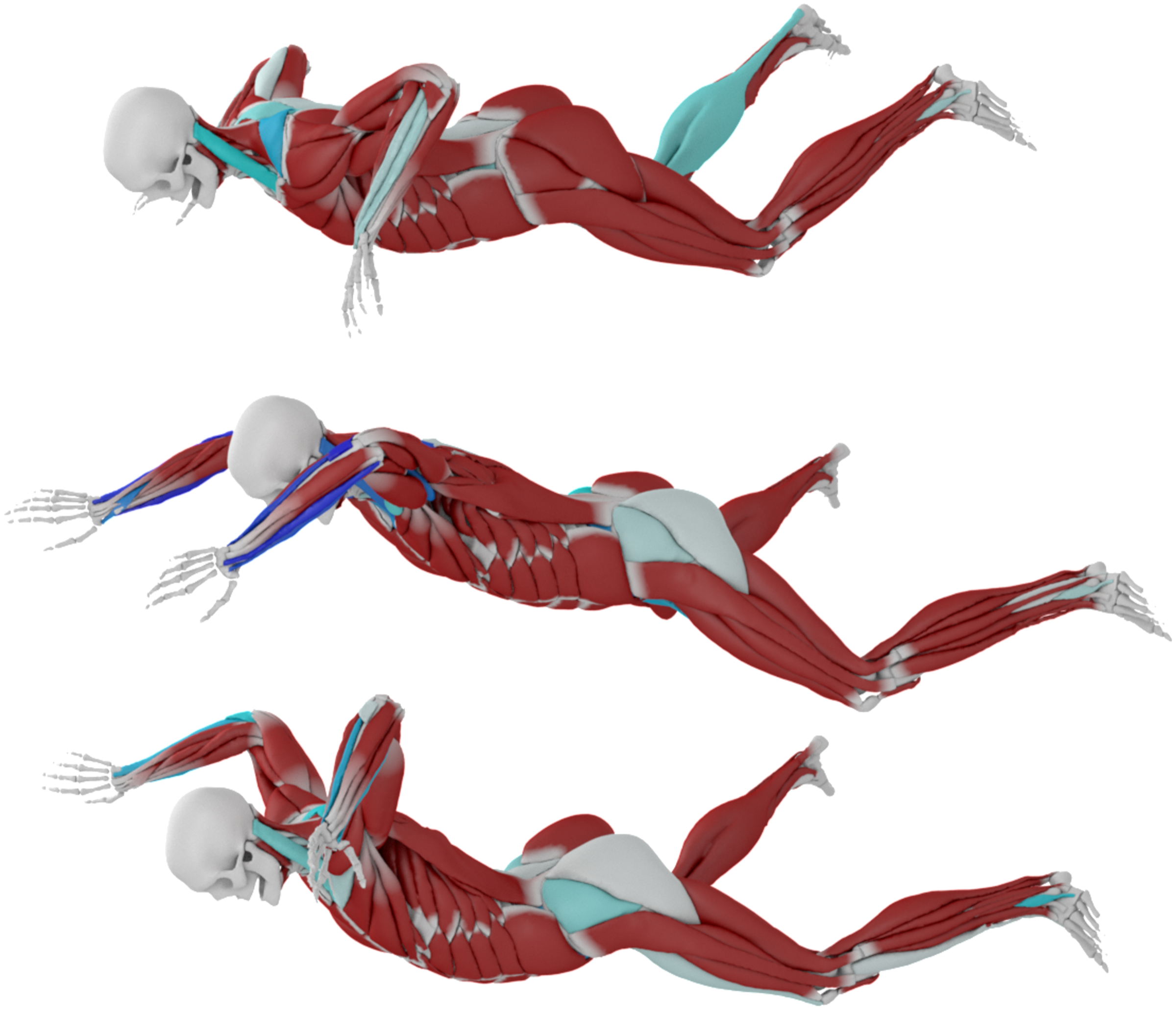}}\hfill
    \subfloat[Squat\label{fig:motion-squat}]{\includegraphics[width=0.2580\textwidth]{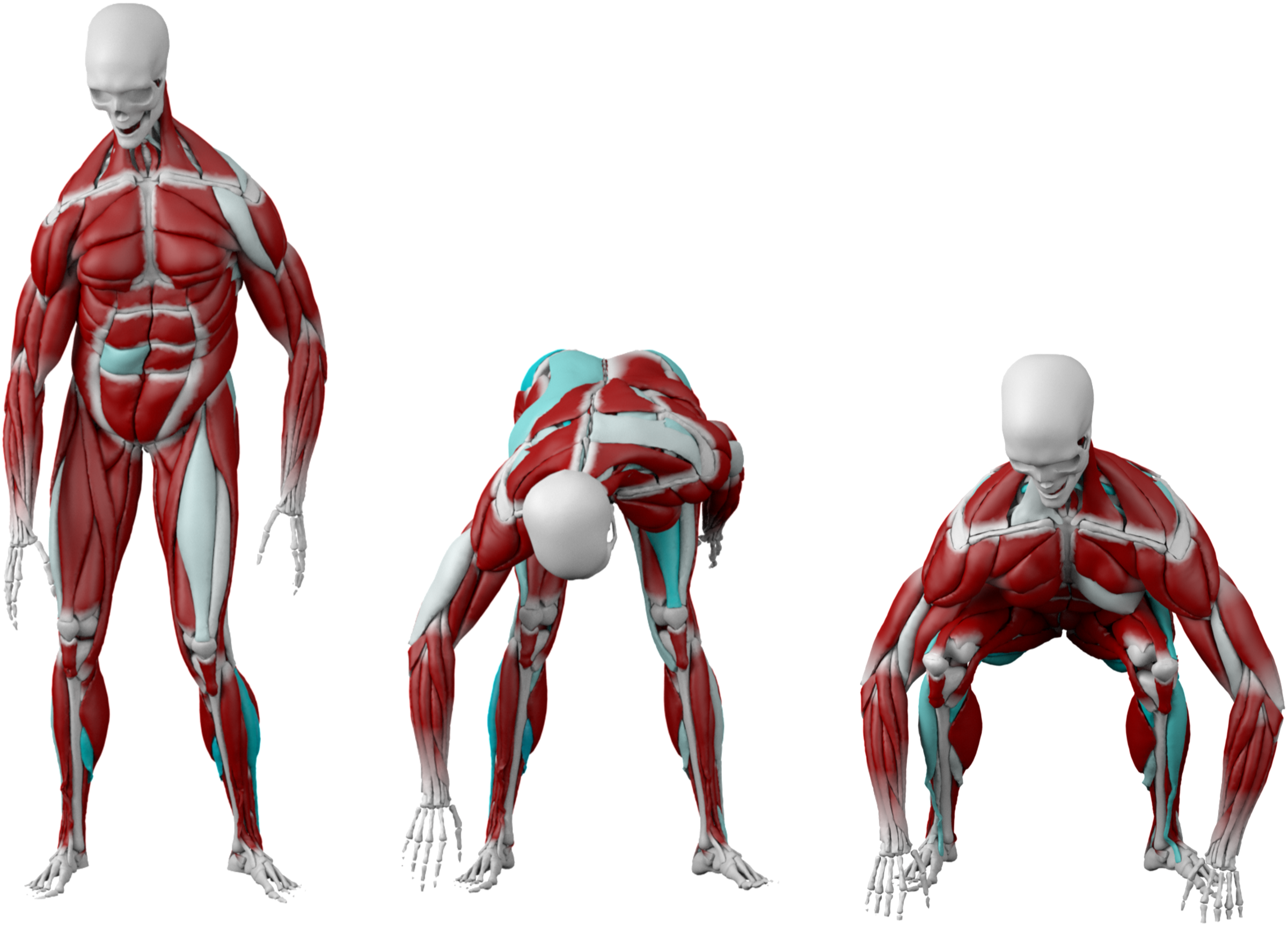}}
    \caption{Volumetric muscle simulation results for four reference motions: (a)~walking, (b)~kicking, (c)~swimming, and (d)~squatting. Blue intensity encodes activation (darker~=~higher); along the red--white axis, whiter is tendon and redder is muscle belly.}
    \label{fig:motions}
\end{figure*}

Figure~\ref{fig:motions} shows our full pipeline applied to four reference-motion clips that span a range of dynamic regimes: a periodic, ground-contact \emph{walk}; a balance-critical \emph{kick} that loads a single support leg; a free-floating \emph{swim} driven by a surrogate fluid model~\cite{vaxenburg2025-fly-mujoco}; and a \emph{squat} that demands whole-body coordination under load. For each clip, the RL stage discovers per-muscle activations that track the reference on the 1D asset, and these activations drive the 194-muscle volumetric body through a single forward XPBD simulation. Each motion trains in under 2 hours, and the walk can be repeated indefinitely if the reference motion is cyclic. Table~\ref{tab:perf} reports the runtime and memory cost of this full-body volumetric simulation on a single RTX~4090. Swimming has no ground contact to react against, so rather than run a full fluid solver we give each body segment an ellipsoidal drag proxy~\cite{vaxenburg2025-fly-mujoco} for the policy to push against.

\begin{figure}[htbp]
    \centering
    \includegraphics[width=\linewidth]{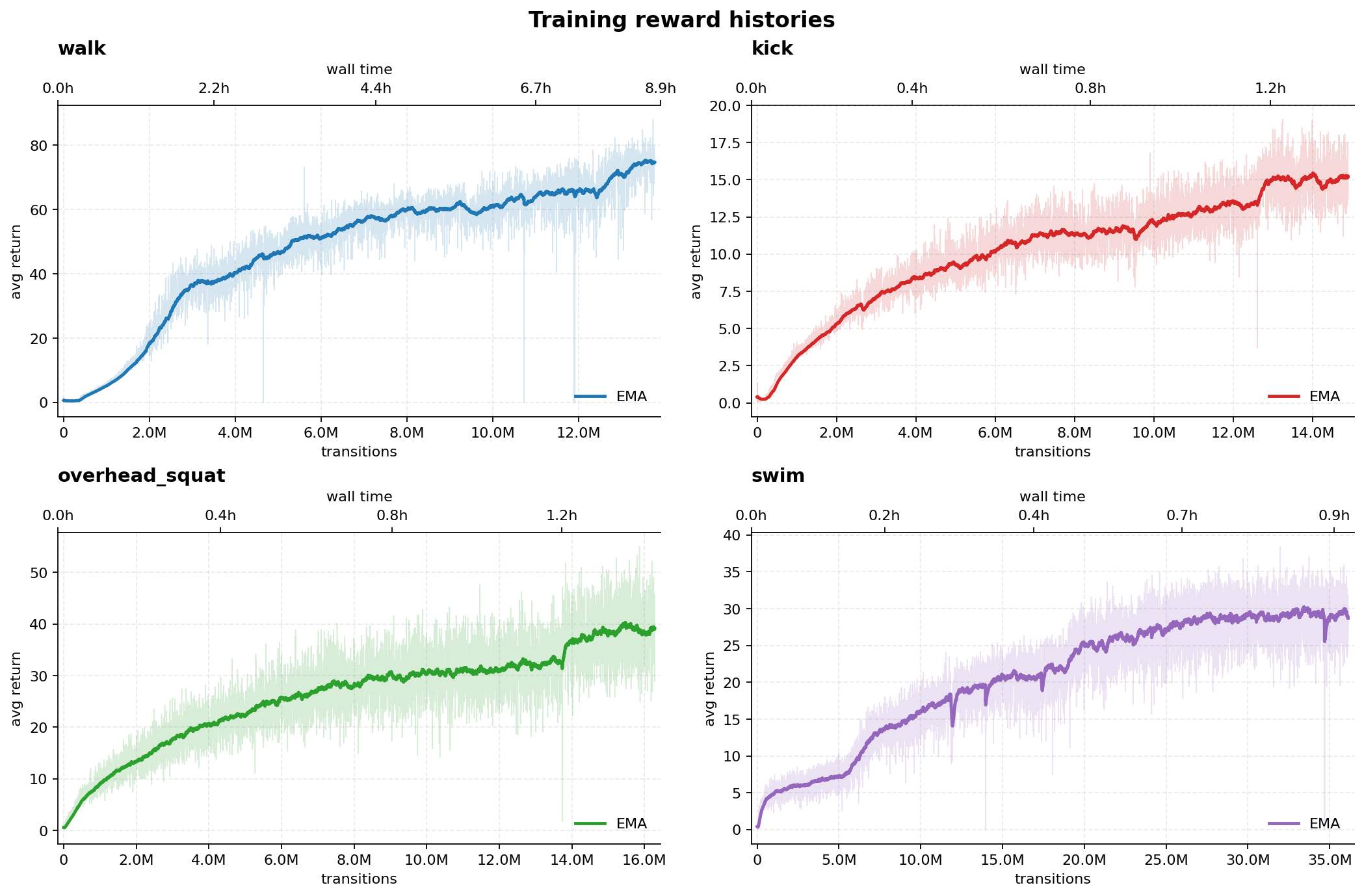}
    \caption{Tracking-reward histories during RL training, one panel per reference motion (each motion trains in under 2 hours). A cyclic reference motion yields an indefinitely repeatable walk.}
    \label{fig:reward_trans}
\end{figure}

\begin{figure}[t]
    \centering
    \includegraphics[width=\linewidth]{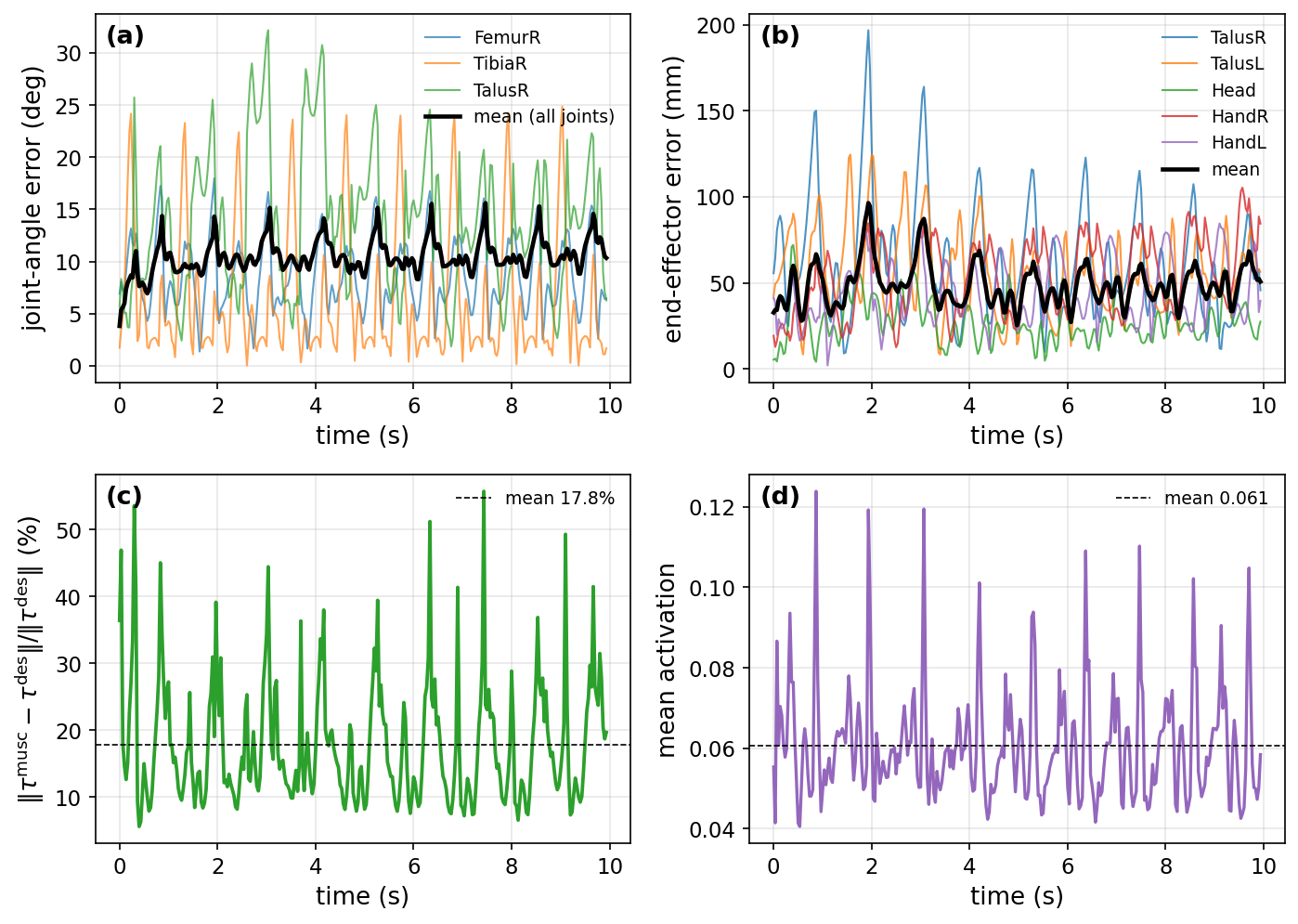}
    \caption{Per-frame evaluation of the trained walking controller over a deterministic $10$\,s rollout: (a)~joint-angle tracking error (Eq.~\ref{eq:reward}, $r_{\text{pose}}$), (b)~end-effector error (Eq.~\ref{eq:reward}, $r_{\text{ee}}$), (c)~relative torque residual (Eq.~\ref{eq:muscle_loss}, $L_{\text{torque}}$), and (d)~mean muscle activation (Eq.~\ref{eq:muscle_loss}, $L_{\text{reg}}$). Dashed lines in (c,\,d) mark the rollout mean.}
    \label{fig:rl-eval}
\end{figure}

Figure~\ref{fig:reward_trans} plots the tracking reward over the course of RL training for all four motions. Every curve rises steadily and plateaus near its maximum, confirming that the co-trained policy and muscle networks (Alg.~\ref{alg:train}) jointly converge to a high-quality tracking solution despite the underdetermined, non-negative, and nonlinear activation-to-torque map. Figure~\ref{fig:rl-eval} quantifies the motion quality of a deterministic rollout of the trained walk ($300$ control steps at $30$\,Hz). Panel~(a) is the joint-angle tracking error, the $SO(3)$ geodesic distance penalized by $r_{\text{pose}}$ in Eq.~\ref{eq:reward}, averaging $10.5^\circ$ over the $18$ tracked joints. Panel~(b) is the COM-aligned position error of the five end effectors behind $r_{\text{ee}}$ in Eq.~\ref{eq:reward}, averaging $49$\,mm. Panel~(c) is the relative torque residual, a scale-free counterpart of $L_{\text{torque}}$ in Eq.~\ref{eq:muscle_loss}, at $17.8\%$ on average. Panel~(d) is the mean activation over all muscles, the same minimum-effort quantity as $L_{\text{reg}}$ in Eq.~\ref{eq:muscle_loss} but without the square, at $0.061$. Together these show that the generated motion tracks the reference closely, yet does not follow it exactly: the activations have to keep the motion physically feasible, which is precisely what we expect.

\begin{table}[t]
    \centering
    \small
    \caption{Runtime of the full-body volumetric simulation over 50 frames. Peak GPU memory excludes the ${\approx}\,386$\,MiB CUDA/Warp context.}
    \label{tab:perf}
    \begin{tabular}{@{}lr@{}}
        \toprule
        Quantity & Value \\
        \midrule
        GPU                       & RTX~4090 \\
        Features                  & collision, multigrid, fiber \\
        Substeps                  & $10$ \\
        Timestep $\Delta t$       & $1/24$\,s \\
        Vertices                  & $172{,}717$ \\
        Tetrahedra                & $468{,}162$ \\
        Simulation (no export)    & $18$\,ms/frame \\
        \quad collision / other   & $11$\,ms / $7$\,ms per frame \\
        Simulation (USD export)   & $32$\,ms/frame \\
        Peak GPU memory           & ${\approx}\,98$\,MiB \\
        \bottomrule
    \end{tabular}
\end{table}

\subsection{Validation Against EMG}
\label{sec:res-emg}

\begin{figure}[htbp]
    \centering
    \includegraphics[width=\linewidth]{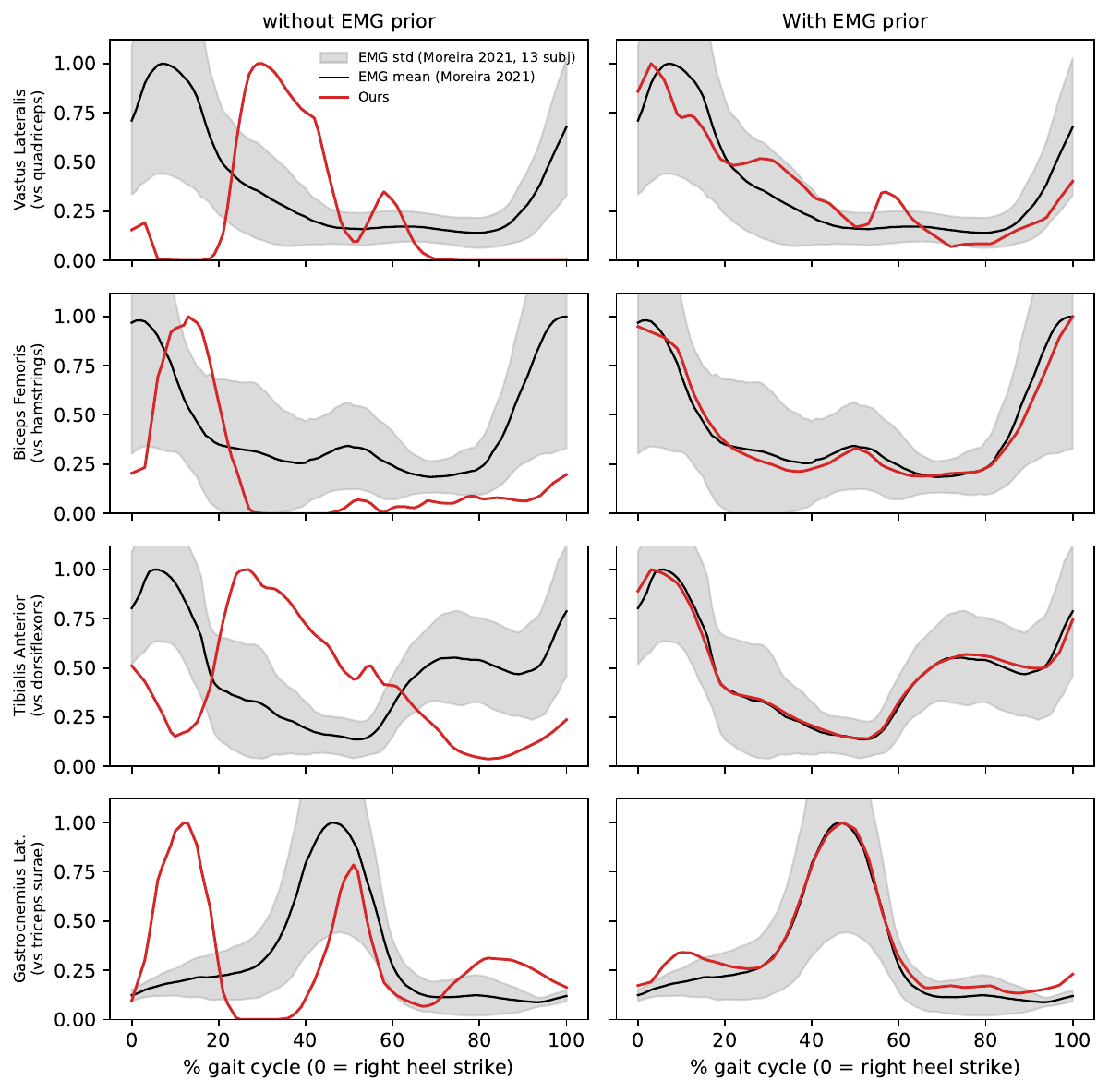}
    \caption{Validation against real walking sEMG~\cite{moreira2021lower} on four right-leg channels (each vs.\ its synergist group), phase-aligned to heel strike~\cite{zeni2008two} and peak-normalized. \emph{Left:} without the prior, our activations (red) leave the $13$-subject human band (gray) over most of the cycle (mean correlation $-0.04$). \emph{Right:} with it, they track the human mean within the band ($0.97$) while the locomotion is unchanged.}
    \label{fig:emg}
\end{figure}

We now test the effect of the EMG prior on physiological plausibility, comparing the recovered activations against the Moreira sEMG~\cite{moreira2021lower} following the synergist-group and gait-alignment protocol of \S\ref{sec:emg-prior} (Fig.~\ref{fig:emg}).

Without the prior, the policy leaves the human $\pm1$\,std band over most of the gait cycle (Fig.~\ref{fig:emg}, left; mean in-band $20\%$, mean correlation $-0.04$), confirming that the redundancy is resolved non-physiologically. Adding it to muscle-network training steers that redundancy toward the measured solution while leaving the policy's motion objective untouched: the recovered activations now track real sEMG (Fig.~\ref{fig:emg}, right; mean in-band $91\%$, mean correlation $0.97$, all on/off timings aligned) and still drive a stable, faithful gait, at negligible cost and without degrading the visual fidelity of the volumetric deformation. Activation--EMG mismatch in tracking-only controllers is widely reported~\cite{li_towards_2026, myosuite, lee_scalable_2019}; our prior offers a simple remedy.

\subsection{Comparison with Baselines}
\label{sec:res-comparison}

We compare against two representative volumetric muscle systems: Houdini's Otis solver~\cite{HoudiniOtis}, the production standard, and the recent neural musculoskeletal skin-deformation model of Han et al.~\cite{han_neural_2024} (Epic). Otis hand-authors activations from tension lines and simulates the motion with Vertex Block Descent~\cite{chen_vertex_2024}; Han et al.\ recover activations by a per-pose quasistatic equilibrium solve and deform an ML skinning model~\cite{Bailey:2018:FDD} trained on XPBD. Ours are learned inside the RL control loop and are exactly the forces that drive the motion (\S\ref{sec:rl}). We compare all three on the same reference pose. We could not include Lee et al.~\cite{lee_dexterous_2018}. Their optimization-based controller re-runs the volumetric solver \emph{inside} the control loop. This has caused repeated non-convergence, which then makes the control learning fail constantly. We therefore exclude it from the baseline methods.

\begin{figure}[htbp]
    \centering
    \includegraphics[height=0.76\linewidth]{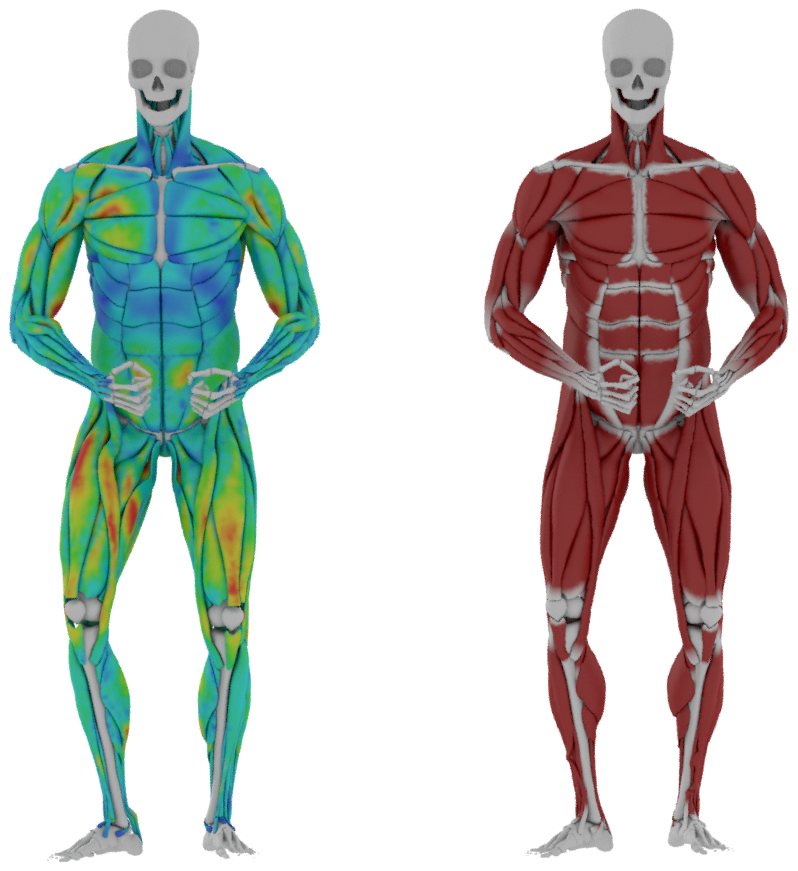}\hspace{2pt}%
    \raisebox{0.14\linewidth}{\includegraphics[height=0.48\linewidth]{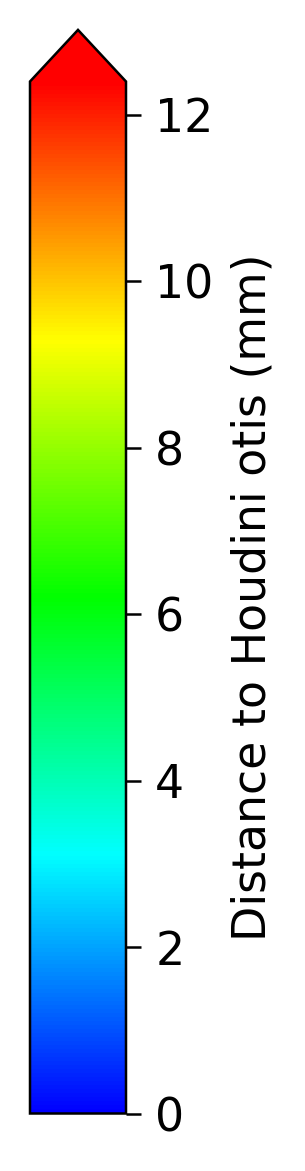}}
    \caption{Comparison against Houdini's Otis solver~\cite{HoudiniOtis} on the same pose. \emph{Left:} our result, colored by per-vertex distance (mm) to the Otis reference (\emph{right}); the distance stays within a few millimeters over most of the body. Our solver runs at $18$\,ms/frame against Otis's $20.8$\,s/frame.}
    \label{fig:otto-vs-houdini}
\end{figure}

\begin{figure}[htbp]
    \centering
    \includegraphics[width=\linewidth]{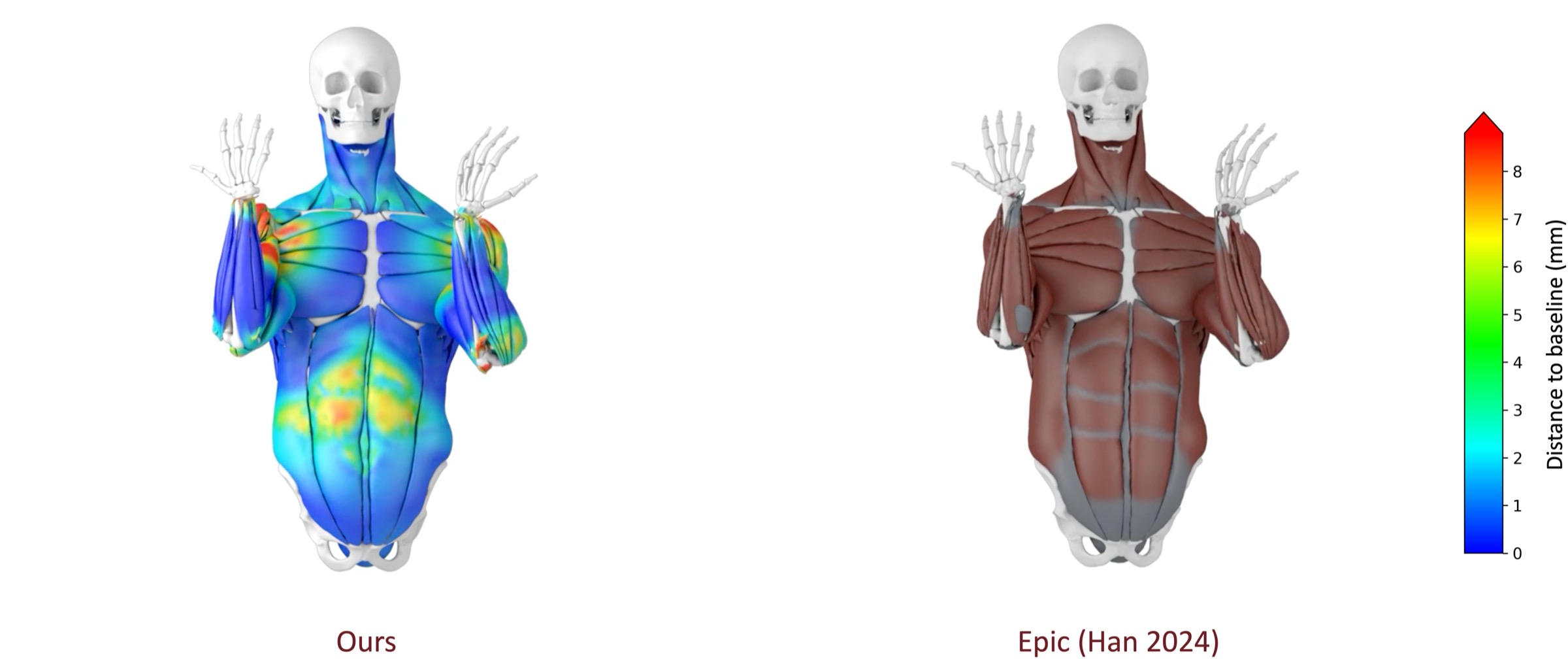}
    \caption{Comparison against the neural musculoskeletal model of Han et al.~\cite{han_neural_2024} (Epic) on the same pose. \emph{Left:} our result, colored by per-vertex distance to their reference (\emph{right}).}
    \label{fig:vs-epic}
\end{figure}

Figure~\ref{fig:otto-vs-houdini} colors our muscle surface (left) by its per-vertex distance to the Otis reference (right) on the same full-body pose. Over most of the body the deviation stays within a few millimeters (blue--green), with larger deviations (yellow--red) at localized regions such as the thighs and shoulders, showing that our physically actuated muscles reproduce production-quality volumetric geometry. The two differ sharply in cost: our GPU solver runs at $18$\,ms/frame (Table~\ref{tab:perf}), whereas Otis, parallelized on the CPU through OpenCL, takes $20.8$\,s/frame on the same AMD Ryzen 9 5950X workstation.
Figure~\ref{fig:vs-epic} repeats this comparison against Han et al.~\cite{han_neural_2024}: the deviation stays small (blue) over most of the torso and grows (yellow--red) at the forearms and mid-abdomen, where the two pipelines differ most in musculature. In both cases the surfaces are close, and the essential difference is provenance: our deformation is produced by the forward-dynamic forces that generate the motion.

\subsection{Ablation Study}
\label{sec:res-ablation}

We ablate each major component of the volumetric pipeline (geometric multigrid, collision handling, and the Hill-type fiber constraint) to isolate its contribution.

\subsubsection{Effect of Multigrid}
\label{sec:abl-mg}
We evaluate our multigrid solver in two complementary settings: a squashed-bunny benchmark, and a full-body volumetric muscle model.

\begin{figure}[htbp]
    \centering
    \includegraphics[width=\linewidth]{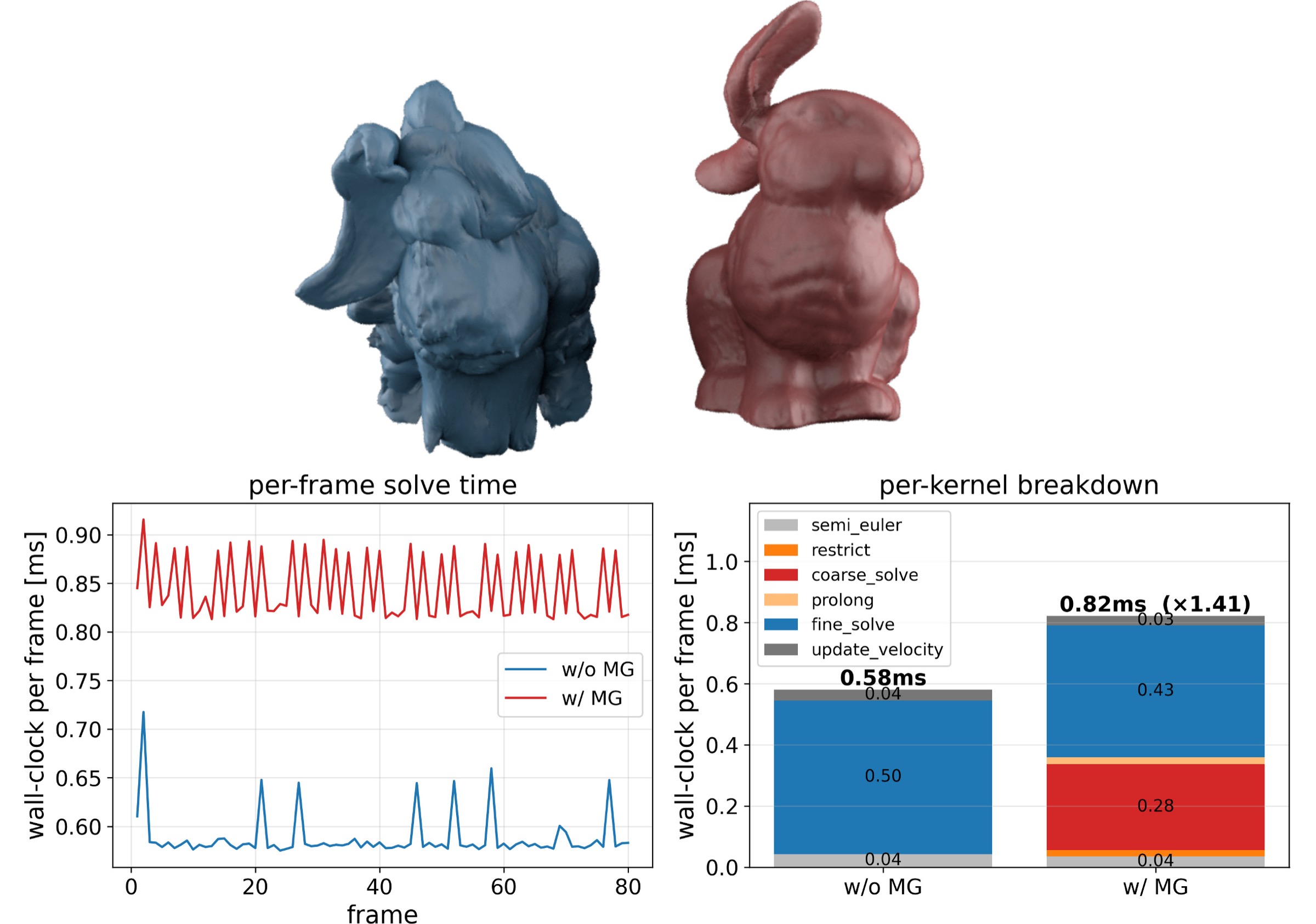}\\[2pt]
    \includegraphics[width=\linewidth]{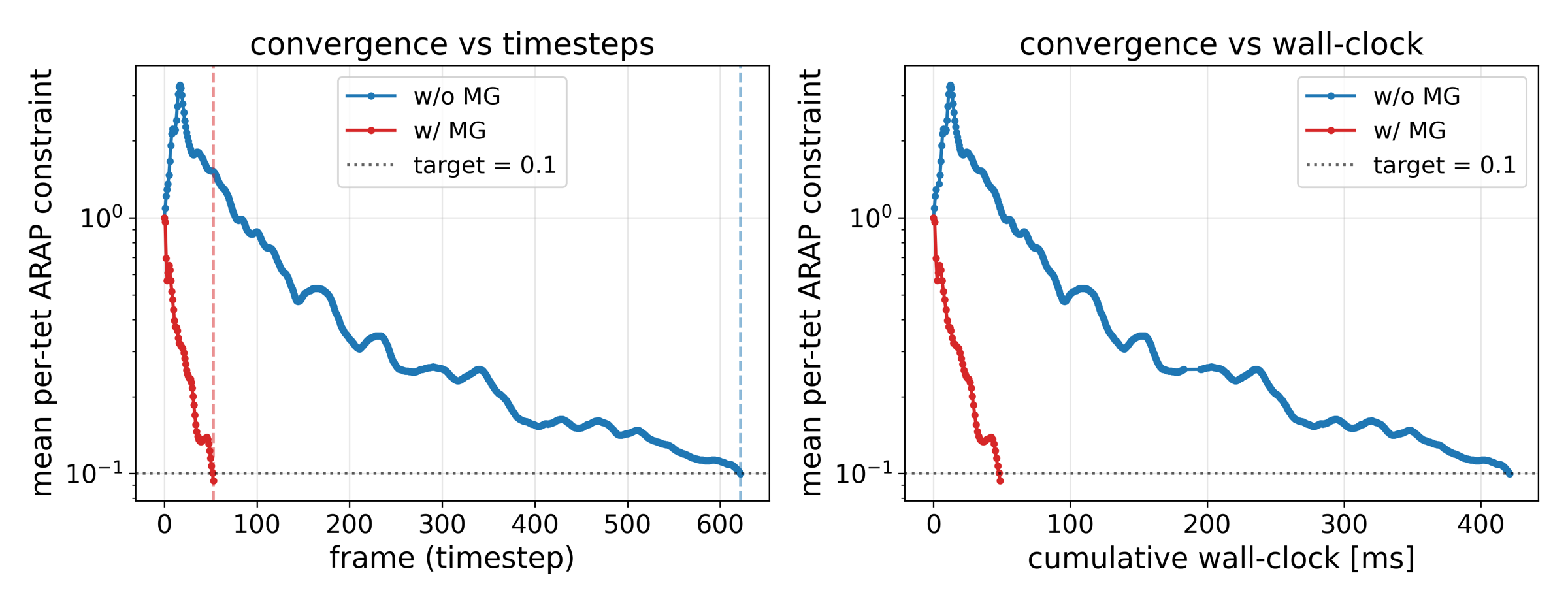}
    \caption{Squashed-bunny benchmark, with (red) vs.\ without (blue) multigrid (MG) under identical parameters. The bunny is squashed, then released. \emph{Top:} recovered shape at the same frame (53). \emph{Middle:} per-frame timing. \emph{Bottom:} residual history $C$ against timesteps and wall-clock. MG converges much faster with a minor per-frame overhead.}
    \label{fig:bunny}
\end{figure}

\textbf{Squashed-bunny benchmark.} The bunny ($850$K fine tetrahedra over a $5$K coarse cage) is compressed and released to bounce back under identical parameters with and without multigrid (MG). Figure~\ref{fig:bunny} shows MG converges faster with better deformation quality, its main cost being the coarse-level solve. Table~\ref{tab:mg} confirms this: MG reaches a fixed tolerance (constraint $C<=0.1$, 2-norm for vector $\mathbf{C}$) in 53 frames and 49.1\,ms vs.\ 621 frames and 418.9\,ms without ($11.7\times$ fewer steps, $8.5\times$ faster), and at 81 frames reaches $C{=}0.082$ where the fine-only solver stalls at $C{=}0.979$. Its small per-frame overhead ($0.84$ vs.\ $0.59$\,ms) is far outweighed by the fewer steps needed.

\begin{figure}[htbp]
    \centering
    \includegraphics[width=\linewidth]{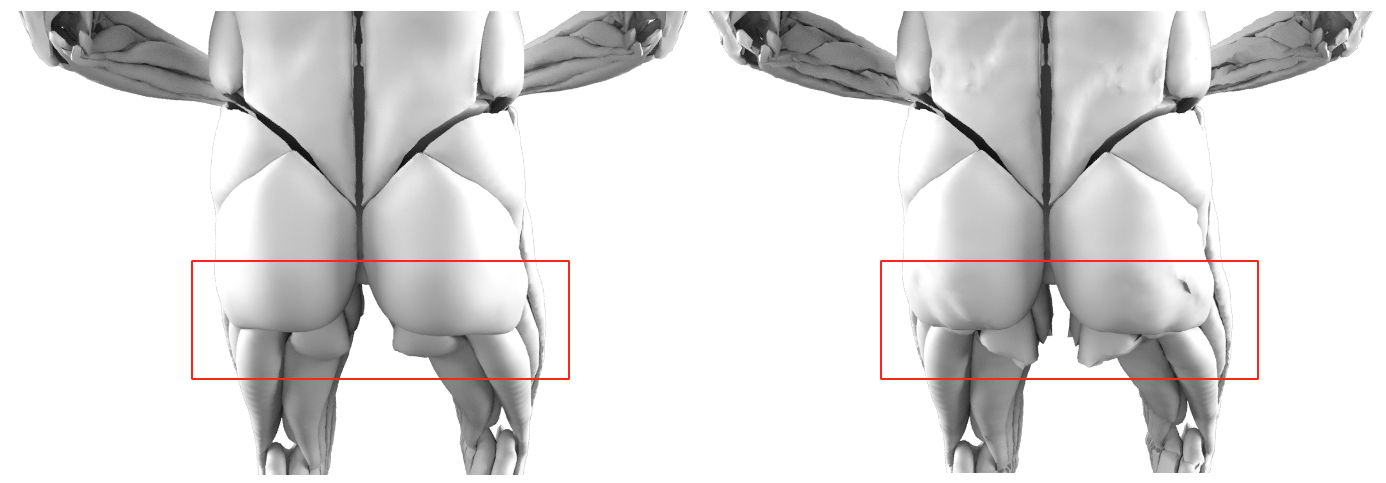}\\[3pt]
    \includegraphics[width=\linewidth]{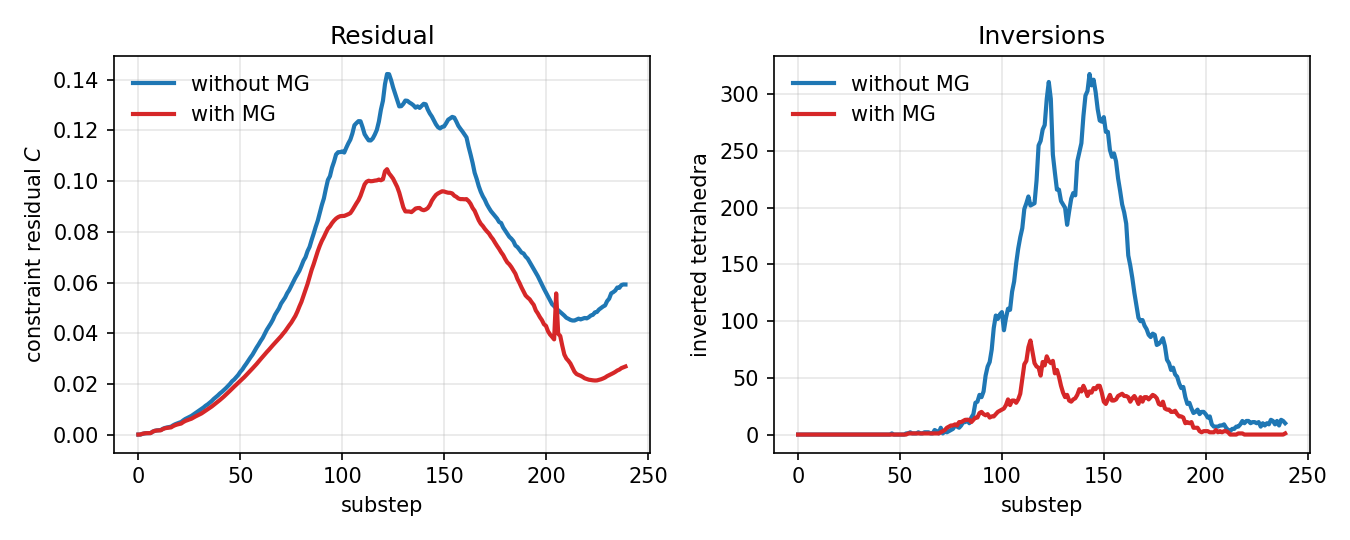}
    \caption{Multigrid (MG) on the full-body volumetric muscle, with (left) and without (right). \emph{Top:} deformation under the high stiffness needed for realistic behavior. \emph{Bottom:} constraint residual $C$ (left) and inverted-tetrahedron count per substep (right).}
    \label{fig:human-mg}
\end{figure}

\textbf{Full-body muscle.} Figure~\ref{fig:human-mg} shows these benefits carry over to the full-body volumetric muscle model. MG is what makes our high-stiffness volumetric simulation both stable and efficient. Figure~\ref{fig:human-mg} (bottom) quantifies this on the same example as Figure~\ref{fig:otto-vs-houdini}: MG holds both the constraint residual and the inverted-tetrahedron count far lower throughout the deformation, roughly a $3.7\times$ reduction in peak inversions.

\begin{table}[htbp]
    \centering
    \small
    \caption{Multigrid (MG) on the squashed-bunny benchmark (Fig.~\ref{fig:bunny}). Reach tolerance $10^{-1}$: $\approx\!11.7\times$ fewer frames and $\approx\!8.5\times$ less wall-clock. At 81 frames: MG $C{=}0.082$ vs without MG $C{=}0.979$.}
    \label{tab:mg}
    \begin{tabular}{@{}lrrrrr@{}}
        \toprule
         & \multicolumn{3}{c}{Reach tolerance} & \multicolumn{2}{c}{At 81 frames} \\
        \cmidrule(lr){2-4} \cmidrule(lr){5-6}
        Solver & Frames & Wall-clock & \shortstack{$C$\\[-1pt]\scriptsize(resid.,\,$\downarrow$)} & Time/frame & \shortstack{$C$\\[-1pt]\scriptsize(resid.,\,$\downarrow$)} \\
        \midrule
        w/ MG  & 53  & 49.1\,ms  & 0.094 & 0.84\,ms & 0.082 \\
        w/o MG & 621 & 418.9\,ms & 0.099 & 0.59\,ms & 0.979 \\
        \bottomrule
    \end{tabular}
\end{table}

\subsubsection{Effect of Collision}
\label{sec:abl-collision}
\begin{figure}[htbp]
    \centering
    \includegraphics[width=\linewidth]{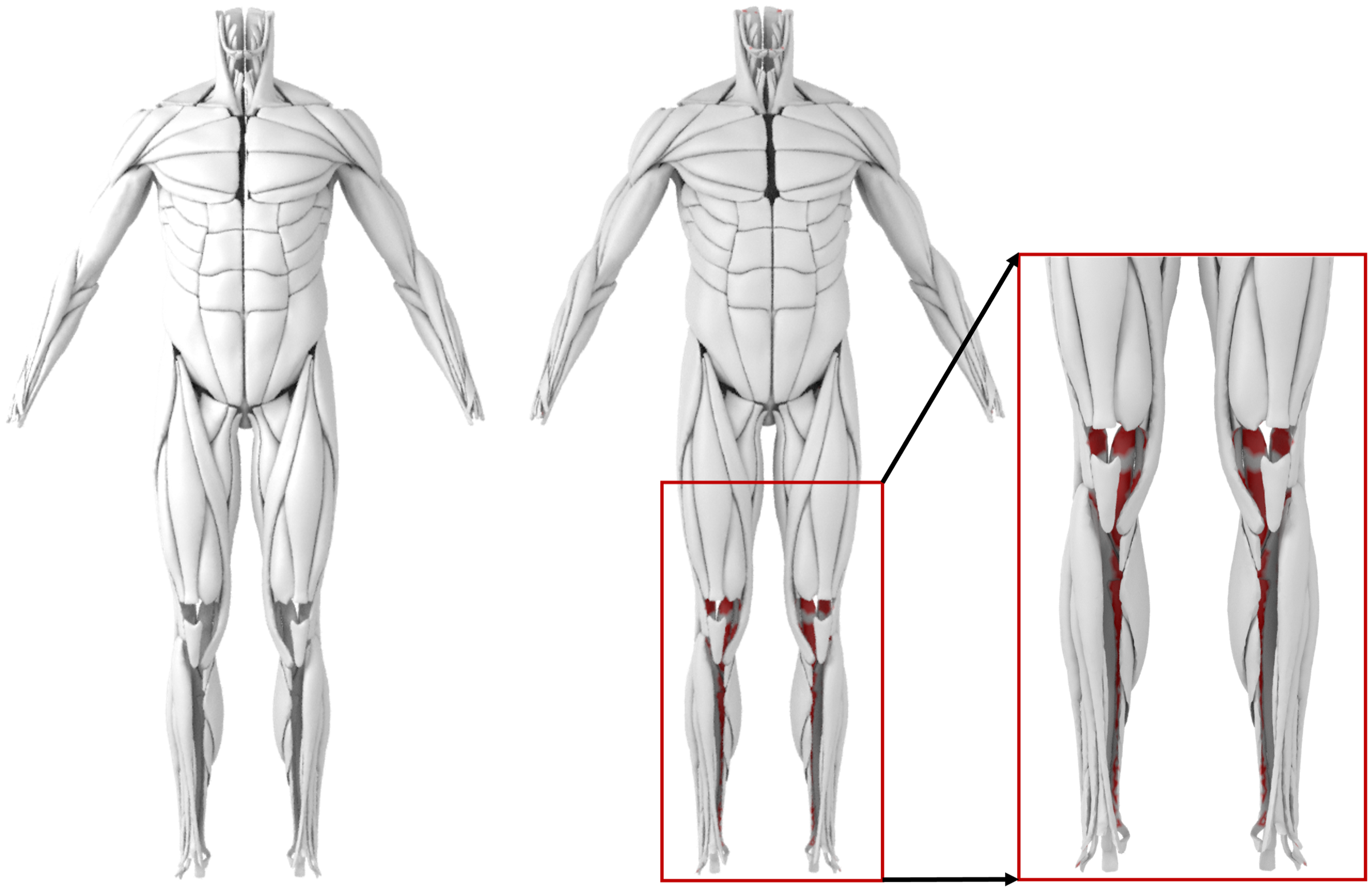}
    \caption{Effect of collision handling, with (\emph{left}) and without (\emph{right}). Red marks the vertices penetrating the bone.}
    \label{fig:coll}
\end{figure}

We ablate the collision handling of \S\ref{sec:collision} by disabling the muscle--bone collision constraints and re-running the same activation trajectory. Figure~\ref{fig:coll} compares the two: with collision handling the muscle bellies slide cleanly over the bones, whereas without it they sink in, concentrated around the knee and lower leg where the muscles wrap most tightly around the skeleton. Enabling collision handling essentially eliminates these interpenetrations.

\subsubsection{Effect of the Fiber Constraint}
\label{sec:abl-fiber}
We isolate the fiber constraint on a biceps brachii held by a single proximal tendon anchor, leaving the belly free to contract as activation ramps from $0$ to $1$. Four constraints are compared under an identical setup: (i)~no fiber force, (ii)~a plain stretch constraint $C=\lVert\mathbf{F}\mathbf{d}\rVert$, (iii)~our Hill--Millard energy constraint (Eq.~\ref{eq:fiber_constraint}), and (iv)~an \emph{anisotropic ARAP} alternative~\cite{kim_dynamic_2020,kim_anisotropic_2019}, built on Kim et al.'s anisotropic invariants and driving the along-fiber stretch toward an activation-set target through a symmetric quadratic well; its invariants, constraint, gradients, and a constitutive comparison against ours are given in the supplementary. Figure~\ref{fig:fiber-actlen} reports the biceps' shortening and girth change at full activation: about $21\%$ shortening for Hill--Millard, $18\%$ for the plain constraint, and $13\%$ for anisotropic ARAP. Hill--Millard reproduces the exact 1D force--length law.

\begin{figure}[htbp]
    \centering
    \includegraphics[width=\linewidth]{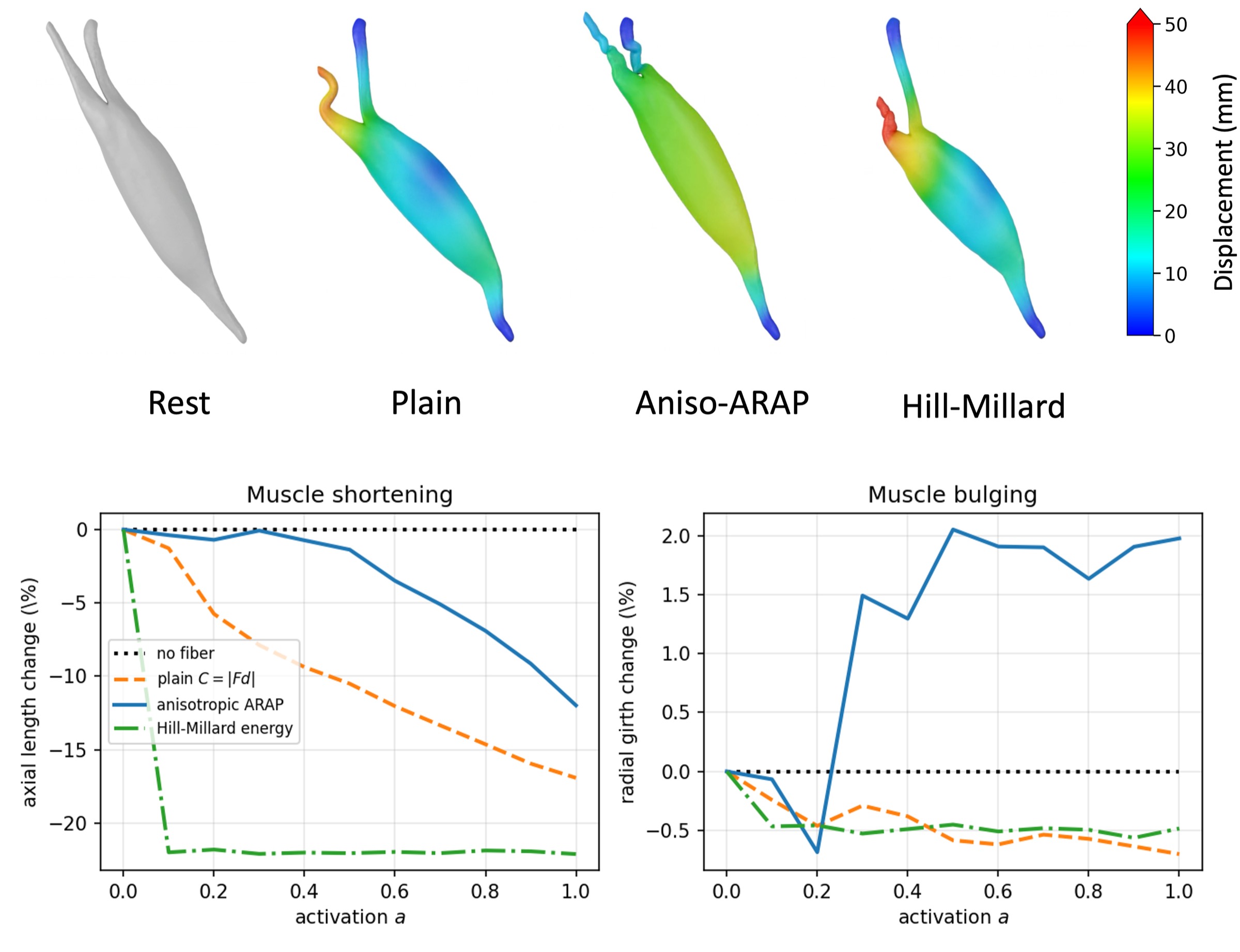}
    \caption{Effect of the four fiber constraints on the biceps. \emph{Top:} deformed biceps at full activation ($a{=}1$), colored by displacement from the no-fiber baseline. \emph{Bottom:} axial length (left) and radial girth (right) vs.\ activation.}
    \label{fig:fiber-actlen}
\end{figure}

\section{Limitations and Future Work}
Ideally, we would incorporate the 3D volumetric muscle simulation into the RL training loop. We tried this integrated pipeline first, but it proved slow, unstable, and uncontrollable for RL training. The difficulty originates from the two-way bone--muscle coupling: spring-based attachments are the naive strategy, but their stiffness has no physiological basis and is hard to tune, since how many springs to use and how stiff they should be are entirely unknown. We therefore use the 1D Hill model as a training surrogate and a separate 3D simulation for evaluation. We hope to achieve ``3D-in-the-loop'' training in future work, which requires two things: a stable and efficient rigid--deformable coupled simulation method, and a physiologically grounded coupling strategy. In addition, while our multigrid performs well on bulky muscles, it is less satisfactory on long, slender muscles, such as some of the muscles near the distal end of the forearm. We attribute this to two factors. First, our geometric coarsening struggles to produce a good coarse mesh for such thin, elongated geometry. Second, because these slender muscles are light and small in volume, their motion is governed more by the boundary conditions (the pin/attachment constraints) than by elastic response, whereas our multigrid is designed to accelerate the elasticity solve. We therefore exclude these slender muscles from the multigrid and solve them at the fine level only.

\section{Conclusion}
We have presented a system that recovers physically consistent muscle activations from a reference motion and turns them into volumetric muscle deformation. Reinforcement learning first explores the muscle activations on a fast 1D Hill-type surrogate, co-training a policy and a muscle network in a single PPO loop and steering the redundant activation solution toward measured human recruitment through an EMG prior. A GPU-parallel XPBD solver then simulates the volumetric muscles under the same Hill-type constitutive law, injected as an energy-based fiber constraint, with a geometric multigrid keeping the stiff solve stable and efficient.

%% file: main.bbl
\begin{thebibliography}{10}
\providecommand{\url}[1]{#1}
\csname url@samestyle\endcsname
\providecommand{\newblock}{\relax}
\providecommand{\bibinfo}[2]{#2}
\providecommand{\BIBentrySTDinterwordspacing}{\spaceskip=0pt\relax}
\providecommand{\BIBentryALTinterwordstretchfactor}{4}
\providecommand{\BIBentryALTinterwordspacing}{\spaceskip=\fontdimen2\font plus
\BIBentryALTinterwordstretchfactor\fontdimen3\font minus
  \fontdimen4\font\relax}
\providecommand{\BIBforeignlanguage}[2]{{%
\expandafter\ifx\csname l@#1\endcsname\relax
\typeout{** WARNING: IEEEtran.bst: No hyphenation pattern has been}%
\typeout{** loaded for the language `#1'. Using the pattern for}%
\typeout{** the default language instead.}%
\else
\language=\csname l@#1\endcsname
\fi
#2}}
\providecommand{\BIBdecl}{\relax}
\BIBdecl
\renewcommand{\BIBentryALTinterwordstretchfactor}{4}

\bibitem{lee_scalable_2019}
S.~Lee, M.~Park, K.~Lee, and J.~Lee, ``\BIBforeignlanguage{en}{Scalable
  muscle-actuated human simulation and control},''
  \emph{\BIBforeignlanguage{en}{ACM Transactions on Graphics}}, vol.~38, no.~4,
  pp. 1--13, Aug. 2019.

\bibitem{myosuite}
V.~Caggiano, H.~Wang, G.~Durandau, M.~Sartori, and V.~Kumar, ``{MyoSuite}: {A}
  contact-rich simulation suite for musculoskeletal motor control,'' in
  \emph{Proceedings of the 4th Annual Learning for Dynamics and Control
  Conference (L4DC)}, ser. Proceedings of Machine Learning Research, vol.
  168.\hskip 1em plus 0.5em minus 0.4em\relax PMLR, 2022, pp. 492--507.

\bibitem{feng_musclevae_2023}
Y.~Feng, X.~Xu, and L.~Liu, ``\BIBforeignlanguage{en}{{MuscleVAE}:
  {Model}-{Based} {Controllers} of {Muscle}-{Actuated} {Characters}},'' Dec.
  2023, arXiv:2312.07340 [cs].

\bibitem{park_magnet_2025}
J.~Park, E.~Jung, J.~Lee, and J.~Won, ``\BIBforeignlanguage{en}{{MAGNET}:
  {Muscle} {Activation} {Generation} {Networks} for {Diverse} {Human}
  {Movement}},'' in \emph{\BIBforeignlanguage{en}{{SIGGRAPH} '25 Conference
  Papers}}.\hskip 1em plus 0.5em minus 0.4em\relax Vancouver BC Canada: ACM,
  Aug. 2025, pp. 1--11.

\bibitem{li_towards_2026}
C.~Li \emph{et~al.}, ``Towards {Embodied} {AI} with {MuscleMimic}: {Unlocking}
  full-body musculoskeletal motor learning at scale,'' Mar. 2026,
  arXiv:2603.25544 [cs].

\bibitem{lee_dexterous_2018}
S.~Lee, R.~Yu, J.~Park, M.~Aanjaneya, E.~Sifakis, and J.~Lee,
  ``\BIBforeignlanguage{en}{Dexterous manipulation and control with volumetric
  muscles},'' \emph{\BIBforeignlanguage{en}{ACM Transactions on Graphics}},
  vol.~37, no.~4, pp. 1--13, Aug. 2018.

\bibitem{schulman_proximal_2017}
J.~Schulman, F.~Wolski, P.~Dhariwal, A.~Radford, and O.~Klimov, ``Proximal
  {Policy} {Optimization} {Algorithms},'' Aug. 2017, arXiv:1707.06347 [cs].

\bibitem{peng_deepmimic_2018}
X.~B. Peng, P.~Abbeel, S.~Levine, and M.~Van De~Panne,
  ``\BIBforeignlanguage{en}{{DeepMimic}: example-guided deep reinforcement
  learning of physics-based character skills},''
  \emph{\BIBforeignlanguage{en}{ACM Transactions on Graphics}}, vol.~37, no.~4,
  pp. 1--14, Aug. 2018.

\bibitem{millard_flexing_2013}
M.~Millard, T.~Uchida, A.~Seth, and S.~L. Delp,
  ``\BIBforeignlanguage{en}{Flexing {Computational} {Muscle}: {Modeling} and
  {Simulation} of {Musculotendon} {Dynamics}},''
  \emph{\BIBforeignlanguage{en}{Journal of Biomechanical Engineering}}, vol.
  135, no.~2, p. 021005, Feb. 2013.

\bibitem{moreira2021lower}
L.~Moreira, J.~Figueiredo, P.~Fonseca, J.~P. Vilas-Boas, and C.~P. Santos,
  ``Lower limb kinematic, kinetic, and {EMG} data from young healthy humans
  during walking at controlled speeds,'' \emph{Scientific Data}, vol.~8, no.~1,
  p. 103, 2021.

\bibitem{Muller2008-HPBD}
M.~M\"{u}ller, ``{Hierarchical Position Based Dynamics},'' in \emph{Workshop in
  Virtual Reality Interactions and Physical Simulation "VRIPHYS" (2008)}.\hskip
  1em plus 0.5em minus 0.4em\relax The Eurographics Association, 2008, pp.
  1--10.

\bibitem{Li2025-MGPBD}
C.~Li \emph{et~al.}, ``{MGPBD}: {A} {Multigrid} {Accelerated} {Global} {XPBD}
  {Solver},'' in \emph{{SIGGRAPH} '25 Conference Papers}.\hskip 1em plus 0.5em
  minus 0.4em\relax ACM, 2025, pp. 111:1--111:11.

\bibitem{muller_position_2007}
M.~M\"uller, B.~Heidelberger, M.~Hennix, and J.~Ratcliff, ``Position based
  dynamics,'' \emph{Journal of Visual Communication and Image Representation},
  vol.~18, no.~2, pp. 109--118, 2007.

\bibitem{physx}
{NVIDIA}, ``{NVIDIA PhysX SDK},'' \url{https://developer.nvidia.com/physx-sdk},
  2024, accessed: 2026-06-08.

\bibitem{Houdini}
{SideFX}, ``Houdini documentation,''
  \url{https://www.sidefx.com/docs/houdini/}, 2023, accessed: 2026-06-08.

\bibitem{macklin_xpbd_2016}
M.~Macklin, M.~Müller, and N.~Chentanez, ``\BIBforeignlanguage{en}{{XPBD}:
  position-based simulation of compliant constrained dynamics},'' in
  \emph{\BIBforeignlanguage{en}{Proceedings of the 9th {International}
  {Conference} on {Motion} in {Games}}}.\hskip 1em plus 0.5em minus 0.4em\relax
  Burlingame California: ACM, Oct. 2016, pp. 49--54.

\bibitem{bender_position-based_2014}
J.~Bender, D.~Koschier, P.~Charrier, and D.~Weber, ``Position-based simulation
  of continuous materials,'' \emph{Computers \& Graphics}, vol.~44, pp. 1--10,
  2014.

\bibitem{Bender2017-Survey-PBD}
J.~Bender, M.~M\"{u}ller, and M.~Macklin, ``A survey on position based
  dynamics, 2017,'' in \emph{Proceedings of the European Association for
  Computer Graphics: Tutorials}, ser. EG '17.\hskip 1em plus 0.5em minus
  0.4em\relax Goslar, DEU: Eurographics Association, 2017.

\bibitem{Briggs2000-MgTut}
W.~L. Briggs, V.~E. Henson, and S.~F. McCormick, \emph{A Multigrid Tutorial},
  2nd~ed.\hskip 1em plus 0.5em minus 0.4em\relax Philadelphia, PA, USA: Society
  for Industrial and Applied Mathematics, 2000.

\bibitem{zhu2010-Multigrid-Solid}
Y.~Zhu, E.~Sifakis, J.~Teran, and A.~Brandt, ``An efficient multigrid method
  for the simulation of high-resolution elastic solids,'' \emph{ACM
  Transactions on Graphics}, vol.~29, no.~2, pp. 16:1--16:18, Apr. 2010.

\bibitem{Mercier2024-Rigidifying}
A.~Mercier-Aubin and P.~G. Kry, ``A {Multi}-layer {Solver} for {XPBD},''
  \emph{Computer Graphics Forum}, vol.~43, no.~8, p. e15186, 2024.

\bibitem{Georgii-Westermann2006-multigrid}
J.~Georgii and R.~Westermann, ``A multigrid framework for real-time simulation
  of deformable bodies,'' \emph{Computers \& Graphics}, vol.~30, no.~3, pp.
  408--415, 2006.

\bibitem{kim_dynamic_2020}
T.~Kim and D.~Eberle, ``\BIBforeignlanguage{en}{Dynamic deformables:
  implementation and production practicalities},'' in
  \emph{\BIBforeignlanguage{en}{{ACM} {SIGGRAPH} 2020 {Courses}}}.\hskip 1em
  plus 0.5em minus 0.4em\relax Virtual Event USA: ACM, Aug. 2020, pp. 1--182.

\bibitem{kim_anisotropic_2019}
T.~Kim, F.~De~Goes, and H.~Iben, ``\BIBforeignlanguage{en}{Anisotropic
  elasticity for inversion-safety and element rehabilitation},''
  \emph{\BIBforeignlanguage{en}{ACM Transactions on Graphics}}, vol.~38, no.~4,
  pp. 1--15, Aug. 2019.

\bibitem{smith_stable_2018}
B.~Smith, F.~D. Goes, and T.~Kim, ``\BIBforeignlanguage{en}{Stable
  {Neo}-{Hookean} {Flesh} {Simulation}},'' \emph{\BIBforeignlanguage{en}{ACM
  Transactions on Graphics}}, vol.~37, no.~2, pp. 1--15, Apr. 2018.

\bibitem{angles_viper_2019}
B.~Angles \emph{et~al.}, ``\BIBforeignlanguage{en-US}{{VIPER}: {Volume}
  {Invariant} {Position}-based {Elastic} {Rods}},'' Jun. 2019, arXiv:1906.05260
  [cs].

\bibitem{dembia_opensim_2020}
C.~L. Dembia, N.~A. Bianco, A.~Falisse, J.~L. Hicks, and S.~L. Delp,
  ``\BIBforeignlanguage{en-US}{{OpenSim} {Moco}: {Musculoskeletal} optimal
  control},'' \emph{\BIBforeignlanguage{en-US}{PLoS Computational Biology}},
  vol.~16, no.~12, p. e1008493, Dec. 2020.

\bibitem{peng_mimicKit_2025}
X.~B. Peng \emph{et~al.}, ``{MimicKit}: {A} {Reinforcement} {Learning}
  {Framework} for {Motion} {Imitation} and {Control},''
  \url{https://github.com/xbpeng/MimicKit}, 2025.

\bibitem{han_neural_2024}
Y.~Han, Y.~Chen, C.~Ong, J.~Chen, J.~Hicks, and J.~Teran,
  ``\BIBforeignlanguage{en}{A {Neural} {Network} {Model} for {Efficient}
  {Musculoskeletal}-{Driven} {Skin} {Deformation}},''
  \emph{\BIBforeignlanguage{en}{ACM Transactions on Graphics}}, vol.~43, no.~4,
  pp. 1--12, Jul. 2024.

\bibitem{Bailey:2018:FDD}
S.~W. Bailey, D.~Otte, P.~Dilorenzo, and J.~F. O'Brien, ``Fast and {Deep}
  {Deformation} {Approximations},'' \emph{ACM Transactions on Graphics},
  vol.~37, no.~4, pp. 119:1--12, Aug. 2018, presented at SIGGRAPH 2018, Los
  Angeles.

\bibitem{tan_stable_2011}
J.~Tan, C.~K. Liu, and G.~Turk, ``Stable {Proportional}-{Derivative}
  {Controllers},'' \emph{IEEE Computer Graphics and Applications}, vol.~31,
  no.~4, pp. 34--44, Jul. 2011.

\bibitem{mujoco}
E.~Todorov, T.~Erez, and Y.~Tassa, ``{MuJoCo}: {A} physics engine for
  model-based control,'' in \emph{2012 {IEEE}/{RSJ} {International}
  {Conference} on {Intelligent} {Robots} and {Systems} ({IROS})}.\hskip 1em
  plus 0.5em minus 0.4em\relax Vilamoura-Algarve, Portugal: IEEE, 2012, pp.
  5026--5033.

\bibitem{mixamo}
{Adobe}, ``{Mixamo},'' \url{https://www.mixamo.com}, 2026, accessed:
  2026-06-08.

\bibitem{anderson_static_2001}
F.~C. Anderson and M.~G. Pandy, ``\BIBforeignlanguage{en}{Static and dynamic
  optimization solutions for gait are practically equivalent},''
  \emph{\BIBforeignlanguage{en}{Journal of Biomechanics}}, vol.~34, no.~2, pp.
  153--161, Feb. 2001.

\bibitem{zeni2008two}
J.~A. Zeni~Jr, J.~G. Richards, and J.~S. Higginson, ``Two simple methods for
  determining gait events during treadmill and overground walking using
  kinematic data,'' \emph{Gait \& Posture}, vol.~27, no.~4, pp. 710--714, 2008.

\bibitem{macklin_snh_2021}
M.~Macklin and M.~M\"{u}ller, ``A {Constraint-based} {Formulation} of {Stable}
  {Neo-Hookean} {Materials},'' in \emph{Proceedings of the 14th ACM SIGGRAPH
  Conference on Motion, Interaction and Games (MIG '21)}.\hskip 1em plus 0.5em
  minus 0.4em\relax ACM, 2021, pp. 12:1--12:7.

\bibitem{irving_invertible_2004}
G.~Irving, J.~Teran, and R.~Fedkiw, ``\BIBforeignlanguage{en}{Invertible finite
  elements for robust simulation of large deformation},'' in
  \emph{\BIBforeignlanguage{en}{Proceedings of the 2004 {ACM}
  {SIGGRAPH}/{Eurographics} symposium on {Computer} animation - {SCA}
  '04}}.\hskip 1em plus 0.5em minus 0.4em\relax Grenoble, France: ACM Press,
  2004, p. 131.

\bibitem{macklin_local_2020}
M.~Macklin, K.~Erleben, M.~M\"uller, N.~Chentanez, S.~Jeschke, and Z.~Corse,
  ``Local {Optimization} for {Robust} {Signed} {Distance} {Field}
  {Collision},'' \emph{Proceedings of the ACM on Computer Graphics and
  Interactive Techniques}, vol.~3, no.~1, pp. 1--17, 2020.

\bibitem{nvidia_warp_2023}
M.~Macklin, ``Warp: {A} {High}-performance {Python} {Framework} for {GPU}
  {Simulation} and {Graphics},'' NVIDIA, \url{https://github.com/nvidia/warp},
  Mar. 2022, nVIDIA GPU Technology Conference (GTC). Accessed: 2026-06-08.

\bibitem{vaxenburg2025-fly-mujoco}
R.~Vaxenburg \emph{et~al.}, ``Whole-body physics simulation of fruit fly
  locomotion,'' \emph{Nature}, vol. 643, no. 8074, pp. 1312--1320, Jul. 2025.

\bibitem{HoudiniOtis}
{SideFX}, ``Otis {Muscle} \& {Tissue} {Sim} geometry node,''
  \url{https://www.sidefx.com/docs/houdini/muscles/muscle_otismuscletissuesim.html},
  2023, accessed: 2026-06-08.

\bibitem{chen_vertex_2024}
A.~H. Chen, Z.~Liu, Y.~Yang, and C.~Yuksel, ``\BIBforeignlanguage{en}{Vertex
  {Block} {Descent}},'' Jun. 2024, arXiv:2403.06321 [cs].

\end{thebibliography}
